\documentclass[12pt]{article}
\usepackage{graphicx}
\usepackage{pdfpages}

\usepackage{amssymb}
\usepackage{amsmath}
\usepackage{cancel}
\usepackage{color}
\usepackage{float}
\usepackage{enumitem}
\usepackage{relsize}
\usepackage[normalem]{ulem}
\usepackage[numbers,sort&compress]{natbib}
\usepackage{hyperref}

\newcommand{\ch}[1]{\textcolor{black}{#1}}
\newcommand{\be}{\begin{equation}}
\newcommand{\ee}{\end{equation}}

\begin{document}
\numberwithin{equation}{section}

\begin{center}
{\Large \bf{ Instant Folded Strings, Dark Energy \\ 
\vspace{3mm}
and a Cyclic Bouncing Universe}}
\\

\vspace{6mm}

\textit{Nissan Itzhaki,${}^{(1)}$ Uri Peleg ${}^{(1)}$ and Paul J. Steinhardt ${}^{(2)}$}
\break \break\break
(1)\  School of Physics and Astronomy, Tel Aviv University, Ramat Aviv, 69978, Israel\\
(2)  Department of Physics, Princeton University, Princeton, NJ 08544, USA
 
\end{center}

\date{\today}
\vspace{5mm}

\begin{abstract}
We present a wholly self-consistent, complete cyclic bouncing cosmology based on components drawn from string theory and constructed in a way that is under perturbative control throughout ({\it e.g.}, with temperature much less than the string scale and string coupling $g_s \ll 1$ at all times).  The cyclic evolution is governed by standard dilaton-gravity in $(3+1)$-dimensions with a perturbatively generated potential and a coupling between the dilaton and a second field that becomes massless at $\phi= \phi_{ESP}$, resulting in an enhanced symmetry point (ESP) that prevents the dilaton from running all the way to zero coupling. 
A central role is played by \emph{instant folded strings} (IFSs) — 
fundamental strings with the unusual property of being much lighter than the string mass while extending far beyond the string length, and violating the Null Energy Condition (NEC).
IFSs are produced classically when the string coupling grows with time, which occurs at two critical points in each cycle. 
In turn, they fulfill a dual function: enabling cosmological bounces and initiating transient epochs of dark-energy domination that naturally transition into slow contraction.
The resulting cosmology eliminates the cosmic singularity and multiverse problems of big bang inflationary models and robustly predicts time-varying IFS-induced dark energy and the absence of primordial B-mode polarization in the cosmic microwave background.

\end{abstract}

\newpage
\baselineskip18pt

\section{Introduction}

Big bang inflationary cosmology was introduced to solve the homogeneity, isotropy and flatness problems of the original big bang model, but what has been discovered over time is that it creates numerous problems of its own, including: ($i$) an {\it entropy problem}  -- how to evolve from a big bang and quantum gravity dominated phase that generate ultra-high entropy to an inflationary phase that requires ultra-low entropy when it starts, a condition that appears to violate the second law of thermodynamics \cite{Penrose:1988mg,Hartle:1983ai,carroll2010,Carroll:2010aj}; 
($ii$) an {\it initial smoothness problem} -- although inflation is supposed to explain the homogeneity and isotropy of the universe, it requires nearly homogeneous and isotropic conditions before it can start \cite{Garfinkle:2023vzf,Ijjas2024}; 
($iii$) an {\it inflaton kinetic energy  problem} -- inflation requires an inflaton field that begins at an unstable position on its effective potential with negligible spatial gradient and kinetic energy much lower than potential energy, when just the opposite conditions are expected after a big bang \cite{Linde:1981mu,Albrecht:1982wi,Steinhardt:1984jj}; ($iv$) {\it multiverse problems} --  
inflation generically leads to eternal inflation  \cite{Steinhardt1983,Vilenkin:1983xq,Linde1986,Linde:1986fd} and a multiverse \cite{Garriga:2005av,Carr2007,Guth:2007ng,Garriga2008,Linde2015,Nomura2017}, rendering the theory  non-predictive and non-falsifiable; ($v$) a {\it $B-$mode polarization problem} -- the same quantum effects that cause inflation to generate 
variations in density also cause inflation to generate tensor variations of the metric \cite{Abbott1984,Starobinsky1983} that produce a distinctive, detectable  “B-mode” polarization pattern in the cosmic microwave background that has not been observed at the expected level \cite{Flauger:2014qra,BICEP2Planck2015,BICEP2021}; and, ($vi$) {\it Transplanckian problems} -- various theoretical arguments imply that there is a limit on the number of $e$-folds of accelerated expansion that are not compatible with inflationary cosmology  \cite{MartinBrandenberger2001,Agrawal:2018own,Palti2019,
Bedroya:2019snp,Brandenberger2021,Bedroya2022,Bedroya2024,Das:2018rpg,Das:2019hto,Brandenberger:2019eni,Brandenberger:2020oav,Vafa2022}.   

Bouncing cosmology \cite{Ijjas:2018qbo,Cook:2020oaj} avoids all these problems by replacing the big bang with a smooth, non-singular transition from a contracting phase to an expanding phase filled with hot matter and radiation and by replacing inflation with slow contraction. The slow contraction phase is a period during which the Hubble parameter is $H(t) \approx p/t$ and the scale factor is $a(t)\propto |t|^p$, where $p < 1/3$ as $t \rightarrow 0^{-}$.  The slow contraction occurs when the energy density of the universe is dominated by a canonical scalar field evolving along a potential whose magnitude is negative.

Avoiding the problems of inflation is then straightforward.
In replacing the bang with a bounce that occurs at energy densities well below the Planck (or species) scale, a quantum gravity dominated phase is avoided, which eliminates the problem of high entropy generation \cite{Ijjas:2020dws,Ijjas:2021zwv}. Smoothing by slow contraction is so robust that it occurs even if the initial deviations from homogeneity and isotropy are large and highly non-perturbative and even if the scalar field responsible for slow contraction has large kinetic energy density \cite{Ijjas:2021gkf,Kist:2022mew,Ijjas2024}.  Slow contraction followed by a bounce and an ordinary radiation-dominated expanding phase entirely avoids inflation and hence the multiverse problem and does not cause the stretching of sub-Planckian modes to super-Hubble scales that leads to a transplanckian problem.  Moreover, bouncing cosmologies avoid the singularity problem that is part of any kind of big bang model.

A range of possible bouncing cosmologies exists. One extreme consists of geodesically complete {\it single bounce models} in which there is a semi-infinite period of contraction connected by a bounce to a semi-infinite period of expansion.  The other extreme is {\it cyclic bouncing models} in which the universe undergoes bounces at regular intervals with periods of decelerated expansion, accelerated expansion (due to dark energy), and slow contraction in between \cite{Steinhardt:2002ih,Ijjas:2019pyf}.  Cyclic models are especially interesting because they make an additional prediction that the dark energy dominating the universe today must \ch{eventually decrease}  with time \cite{Ijjas:2021zwv,Montefalcone:2020vlu,Andrei:2022rhi} such that the current accelerated expansion phase eventually transitions to a slow contraction phase followed by a future bounce.   This prediction is being tested in ongoing and near-future astrophysical observations \cite{DESI:2024mwx,DESI:2025fii}. 

An essential element of all these cosmologies is the bounce.
Several proposals for the bounce have been put forward in the context of effective field theory and string theory \cite{Brandenberger:1988aj,Khoury:2001wf,Khoury:2001bz,Lehners:2011kr,Easson:2011zy,Xue2013,Qiu:2013eoa,Battarra:2014tga,Alexander:2014uaa,Ijjas:2016tpn,Ijjas:2016vtq,Ijjas:2017pei,Farnsworth:2017wzr,Brandenberger:2016vhg,Tukhashvili:2023itb}.  Still, it is fair to say that the precise nature of the bounce is unsettled, particularly how to violate the null energy condition in a consistent and controlled manner.  
Recently, in \cite{InstantCosmology}, it was suggested that Instant Folded Strings (IFSs) \cite{Itzhaki:2018glf} — non-standard strings nucleated classically in certain time-dependent backgrounds in string theory, which exhibit the seemingly contradictory properties of being much lighter than the string mass yet much larger than the string length — can mediate a cosmological bounce.
In this paper, we demonstrate by explicit construction that cyclic universe models consistent with current cosmological observations can be constructed with the help of IFSs combined with other ingredients believed to be present in the string landscape -- namely dilaton-gravity governed by a perturbative potential and an enhanced symmetry point (ESP).  An advantage of an IFS-mediated bounce is that it can be described in a manner that is precise and under sound theoretical control, and be incorporated in a cyclic scenario that is under perturbative control at all stages.

 Given the novel components,  multiple phases and transitions involved, the scenario we propose may at first seem quite complex, especially for readers unfamiliar with some of its elements, such as the IFSs and the ESP. To some extent, complexity is inevitable: any {\it complete} cosmological model must include a sequence of multiple stages to match observations and its description must include the underlying physics at each stage and the mechanism that causes one stage to transition to the next. Our framework is no exception. However, we would like to point out that once these ingredients, both novel and familiar, are put together, the sequence of stages flows rather smoothly and naturally from one to the next, resulting in a cyclic cosmic engine that seems capable of running indefinitely, which is quite remarkable.

The paper is organized as follows. The next section provides a qualitative overview of the IFS-mediated cyclic universe. The rest of the paper is devoted to a more detailed discussion of the scenario. For completeness we start in section 3 with a review of some basic properties of IFSs that are relevant for cosmology.  Sections~4 through~7 describe in detail key stages of a cycle: the ESP, the bounce and reheating, the IFS-induced dark energy phase, and the slow contraction phase, respectively. Section 8 summarizes the full scenario and closes with a plot of the evolution of the Hubble parameter $H(t)$ and the scale factor $a(t)$ for a fully-worked example that is under theoretical control ({\it e.g.,} with string coupling $g_s \ll 1$ throughout the cycle and reheat temperature $T_{\rm rh}$ below the string scale). 
The Appendix presents the quantitative constraints the model parameters must satisfy and the specific choices made for the worked example in Fig.~13.

\section{The scenario in a nutshell}
\label{nutshell}

 As noted in the Introduction, because we are describing a {\it complete} model of the universe that includes novel components that each need to be explained, and, as with any complete cosmological scenario,  involves many stages and transitions, even a broad overview like the one in this section can seem complicated. It's a summary in a nutshell, but for a very big nut! With increasing familiarity, however, we hope the reader will appreciate how the stages flow smoothly and stably from one to the next and combine into a regularly repeating cosmology while satisfying the many constraints.

\begin{figure}[t!]
\includegraphics[width=16cm]{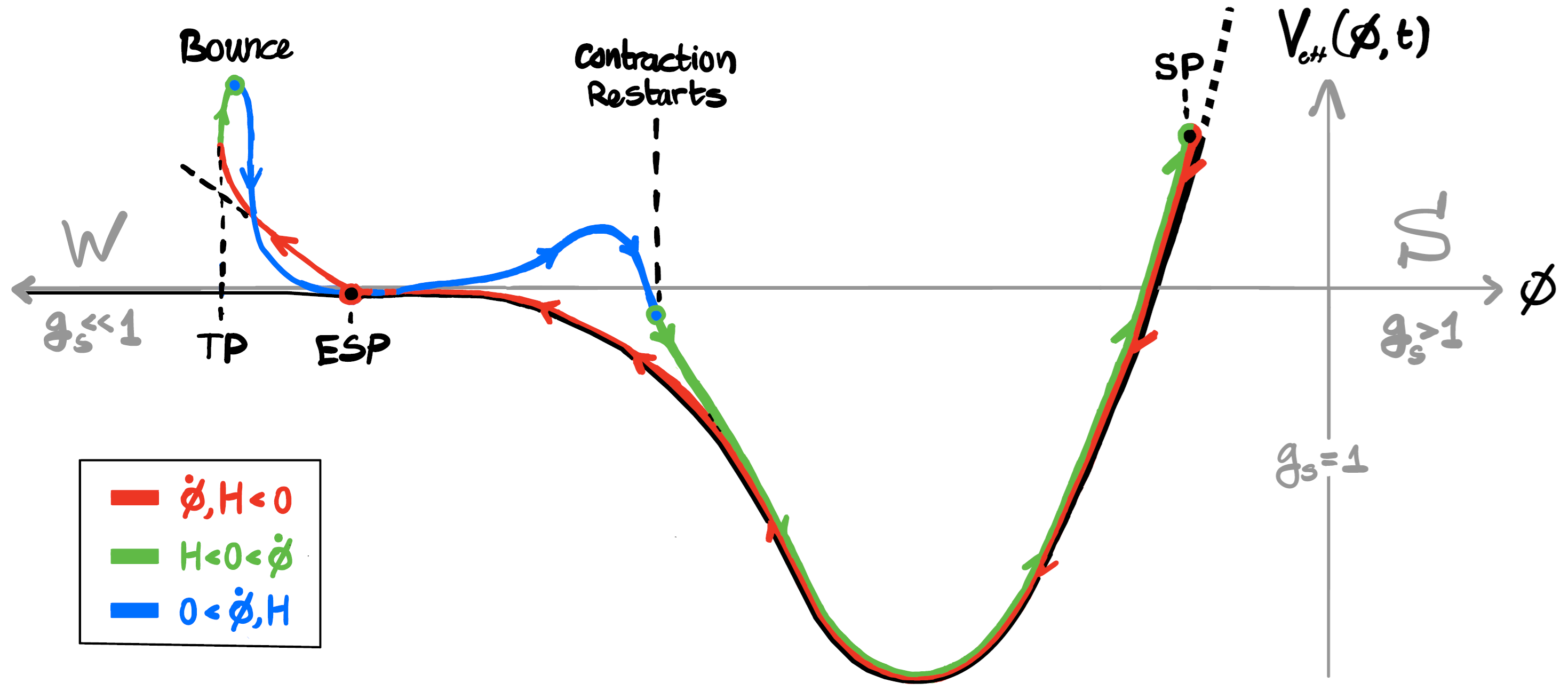}
\centering
\caption{A plot of the effective potential $V_{eff}$, the sum of the underlying dilaton potential $V$, the ESP-induced potential $V_\chi$, and the contribution from the IFS sector. 
To the left is the weak coupling limit (marked $W$) and to the right is the strong coupling limit (marked $S$).
The colors and arrows are used to indicate the direction in which $\dot{\phi}$ is evolving, and whether the universe is expanding or contracting at that time. Due to the time dependence of $V_\chi$, and since IFS production occurs only when $\dot{\phi}>0$, the effective potential is a double-valued function of $\phi$ --- depending on $\dot{\phi}$'s direction --- and traces a self-intersecting trajectory.}
\label{fig:simplecyclic}
\end{figure}

The cyclic universe scenario introduced in this paper is a dilaton-driven cosmology, described using the familiar 4D dilaton-gravity equations, accompanied by the following three ingredients:
\begin{enumerate}

\item An enhanced symmetry point \cite{ESP} - a point, $\phi_{ESP}$,  where some other field, $\chi$, becomes massless.
\item Instant folded strings - non-standard strings that are nucleated classically only when the string coupling grows with time.

\item A potential for the dilaton, $V(\phi)$. We assume, as is typically the case, that the dilaton acquires a potential in perturbation theory,
which in the Einstein frame takes the form
\be
V(\phi)=\sum_{j=1}V_j=\sum_{j=1}c_j g_s^{2+2j}, \qquad g_s\equiv e^{\frac{\kappa\phi}{\sqrt{2}}}.
\ee 
In line with \cite{Dine:1985he} $V(\phi)$ vanishes in the weak coupling limit, $\phi \to -\infty$. As discussed in Section 6, our scenario requires certain adjustments of the potential parameters: $c_2$ has to be negative and large compared to $c_1$, and some higher-order terms have to be positive in order that the potential be bounded below ({\it i.e.}, has an absolute minimum), as illustrated by the black curve in Fig.~\ref{fig:simplecyclic}.

\end{enumerate}

\begin{figure}
\includegraphics[width=15cm]{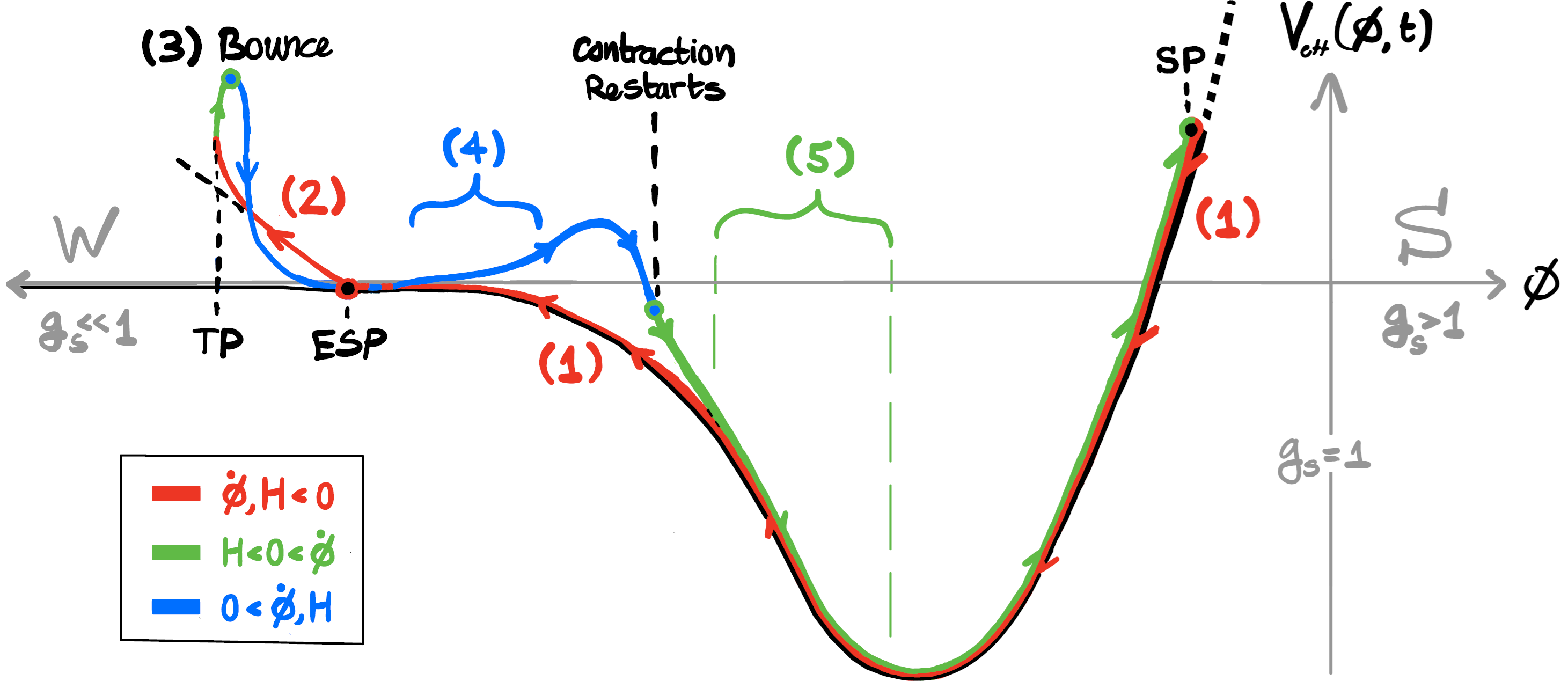}
\centering
\caption{The effective potential $V_{eff}$ throughout the different cosmological epochs. The equations of motion that govern the cyclic evolution are given in equations (\ref{a}) to (\ref{c}) near the end of Section~3. Below is a table that briefly describes each stage and transition while keeping track of the sign changes of $H$, $\dot{H}$, and $\dot{\phi}$.}
\begin{center}
\begin{tabular}{ ||p{1.2cm}||p{0.3cm}||p{0.3cm}||p{0.3cm}||p{12cm}||  }
 \hline
 \multicolumn{5}{||c||}{\textbf{Summary of Each Stage and Transition}} \\
 \hline
 \textbf{Stage}: & $\dot{\phi}$ & $H$ & $\dot{H}$ & \textbf{Description:}\\
 \hline
 ~~\textbf{(1)} & \textbf{--}    & \textbf{--} & \textbf{--} &   Universe slowly contracts as dilaton rolls towards weak coupling\\ 
 \hline
 ~~\textbf{(--)} & \textbf{--} & \textbf{--}&  \textbf{--} & Dilaton crosses the ESP, leading to production of $\chi$ particles\\
 \hline
 ~~\textbf{(2)} & \textbf{--}  & \textbf{--}&  \textbf{--}   & Potential induced by $\chi$ particles exerts a force slowing  the dilaton
 \\
 \hline
 ~~\textbf{(--)} & 0  & \textbf{--}&  \textbf{--}   & Dilaton reaches turning point TP where $\dot{\phi}=0$ \\
 \hline
 ~~\textbf{(3)} & + & \textbf{--} + + & +   + \textbf{--}& A full cosmological bounce: $\dot{\phi}>0$ produces IFSs that induce $\cancel{\mathrm{NEC}}$, reversing the sign of $\dot{H}$, and then $H$; this results in Hubble friction that ends $\cancel{\mathrm{NEC}}$, reversing the sign of $\dot{H}$ again  
 \\
 \hline
 ~~\textbf{(--)} & + & + &  \textbf{--}&   $\chi$ particles decay, reheating the universe; $\dot{\phi}>0$ increases, producing  new IFSs\\
 \hline
 ~~\textbf{(4)} & + & +&  \textbf{--}&  IFS induced friction results in a dark energy-dominated phase that includes today \\
 \hline
 ~~\textbf{(--)} & + & + \textbf{--}  &  \textbf{--} \textbf{--}&  The effective potential becomes negative as the effect of IFSs weakens, causing $H$ to change sign, initiating slow contraction \\
 \hline
 ~~\textbf{(5)} & +  & \textbf{--} &  \textbf{--}& Slow contraction smooths the universe\\
 \hline
 ~~\textbf{(--)} & 0  & \textbf{--} &  \textbf{--} & The steep positive slope of the potential stops the dilaton at SP and a new cycle begins\\
 \hline
\end{tabular}
\end{center}
\label{fig:cyclic}
\end{figure}

Our cyclic scenario is summarized in Fig.~\ref{fig:simplecyclic} (and in Fig.~\ref{fig:cyclic} in greater detail). During each cycle, the dilaton travels from strong to weak coupling and back again.
A good starting point to describe the cycle is when the dilaton is at the point marked by SP in Fig. \ref{fig:simplecyclic}, where $V>0$. This is a turning point where the dilaton reaches its largest value and begins its traversal towards weak coupling. We assume that at SP the universe is contracting, $H<0$, and that the dilaton is at rest, $\dot{\phi}=0$. 
Due to the slope of the potential, the dilaton is pushed towards weak coupling,
$\dot{\phi}<0$, and since $H<0$, the Hubble ``anti-friction" term $3 H \dot{\phi}$  increases the velocity of the dilaton towards $\phi \rightarrow -\infty$.
Since $V(\phi \rightarrow -\infty)=0$, the contracting universe becomes dominated by the kinetic energy density of the dilaton ({\it i.e.}, a period of kination with growing kinetic energy density ensues). 

To avoid a singularity, we design the model so that, as the dilaton moves toward the weak coupling limit at $\phi \rightarrow - \infty$, it encounters an Enhanced Symmetry Point (ESP), as described in \cite{ESP}.  An enhanced symmetry point occurs when  a second field $\chi$, a fermion for example,  is coupled to the dilaton through an interaction of the form \be\label{f} \lambda_{\chi}(\phi-\phi_{ESP})\bar{\chi}\chi\ee
with the property that $\chi$ becomes massless at $\phi_{ESP}$.
As the dilaton $\phi$ evolves towards weak coupling, $\chi$ particles are created quantum mechanically, but  
the creation rate is highly suppressed when $\phi$ is far from $\phi_{ESP}$ because the $\chi$ particles are massive.  However, as the dilaton crosses the ESP,  the $\chi$ particles become massless and their creation rate increases considerably \cite{ESP}. 

Since their mass depends on the dilaton, the creation of the $\chi$ particles affects the evolution of the dilaton in a significant way: it generates a time-dependent potential
\be\label{ft}
V_{\chi}(\phi) = \lambda_{\chi} n_\chi(t) |\phi-\phi_{ESP}|,\ee
where $n_{\chi}$ is the number density of the $\chi$ particles. 
The time dependence of $V_{\chi}(\phi)$ stems from the fact that $n_{\chi} \sim 1/a^3(t)$. 
As is clear from (\ref{ft}), this potential tends to slow down the dilaton and pushes it back towards the ESP \cite{ESP}. 

The physics associated with it was studied extensively in the case of an expanding universe. In our scenario, however, the dilaton crosses the ESP when the universe is contracting.  This raises two novel issues. First, the dilaton has to stop before the universe reaches the singularity. 
The second issue is more subtle. As the dilaton crosses the ESP, the time-independent potential, $V(\phi)$, is negligible compared to $V_{\chi}(\phi)$ and the kinetic energy of the dilaton. Therefore, the dynamics, in particular, the stopping of the dilaton, is determined by the competition between the kinetic energy and $V_{\chi}(\phi)$, which scale differently with $a(t)$ - roughly, $V_{\chi}(\phi)\sim 1/a^3(t) $ and $\dot{\phi}^2\sim 1/a^6(t)$. 
This implies that in a contracting universe, $V_{\chi}(\phi)$ has a limited time to flip the sign of $\dot{\phi}$ before the kinetic energy dominates it. This leads to   \ch{ an upper bound} on the speed at which the dilaton crosses the ESP, $\dot{\phi}_{ESP}$,  which is discussed in detail in  Section \ref{crossing}. There, we show that the conditions needed to flip the sign of $\dot{\phi}$ are easily satisfied provided the reheating temperature  $T_{rh} \lesssim 10^{16}\mbox{GeV}$. 

After the ESP flips the sign of $\dot{\phi}$, the universe is still contracting, but the string coupling $g_s$ is now growing with time.  As a result, IFSs are created. The way the IFS sector affects the dilaton gravity equations of motion is reviewed in detail in Section~\ref{Instant folded strings}. The upshot is that they induce a negative pressure and an extra friction term (on top of the Hubble friction) that roughly takes the form
\be\label{rt}
p_{IFS}\sim -\dot{\phi}^2/g_s^2~~{\rm and}~~~\mbox{IFS friction} \sim \dot{\phi}^2/g_s.
\ee
This \textbf{negative pressure} is a remarkable feature of IFSs that plays a central role in our scenario: namely, it leads to a violation of the NEC that results in a smooth \textit{cosmological bounce}.  More precisely, during a contracting phase in which both $H<0$ and $\dot{H}<0$, IFS production and NEC violation are triggered when $\phi$ reaches a turning point at which $\dot{\phi}$ flips sign from negative to positive. This initiates a smooth continuous bounce in which: ($i$) $\dot{H}$ flips sign from negative to positive, 
($ii$) which causes  $H$ to increase and eventually cross from negative to positive, ($iii$) which eventually causes $\dot{H}$ to become negative once again.  At which point the universe has smoothly transitioned to a standard FRW expanding phase.

In Section \ref{bounce}, we examine the details of the cosmological bounce. In this overview section, we give a rather heuristic argument why 
the IFSs are so efficient in mediating a bounce based on the Bianchi identity.  
During the IFS-dominated phase, at least for a short period, the NEC is violated. According to the Bianchi identity, once the universe is dominated by IFSs and their decay products, we have the effective relation 
\be 
\dot{\rho}_{IFS}+3H(\rho_{IFS}+p_{IFS})=0, \label{Bia}
\ee
where $\rho_{IFS}$ and $p_{IFS}$ represent the energy density and pressure associated with the IFS sector.  Since $p_{IFS}$ is negative and large, the Bianchi identity implies that, in a contracting universe, $\rho_{IFS}$ is decreasing and, as a result, the NEC violation induced by the IFS sector is sufficiently amplified to trigger a cosmological bounce. 
In other words, once a contracting universe enters an NEC-violating phase, the NEC violation effect becomes increasingly strong. This continues until $H$ switches sign and becomes positive, at which stage the Bianchi identity implies that the NEC violation is suppressed.  
In Section 5, we show that this rough argument is valid:  a bounce occurs rather quickly once $\dot{\phi}$ flips its sign.  Moreover, we show that the bounce does not involve a large curvature.
That is, the bounce is entirely described by effective equations of motion that are under good perturbative control.  

The kinetic energy of the dilaton is transferred to the $\chi$-induced potential $V_\chi$, via the action of the ESP mechanism, and then transferred to the IFS sector when $\dot{\phi}>0$ , due to the slope of the $\chi$-induced potential. In this way, the universe is reheated through the eventual decay of the $\chi$-particles and the IFSs into standard model particles. The decay process of IFSs is discussed in Section~\ref{Instant folded strings}, and in more detail in \cite{InstantCosmology}. The $\chi$-particles become progressively diluted as the universe expands and as they decay throughout the reheating process. As a result, the dominance of the $\chi$-induced potential and the IFS sector eventually comes to an end. Then follows a standard period of radiation domination, eventually giving rise to matter domination as the universe cools.

It is natural to suspect that by this stage, the role of the IFSs in our scenario is over for the rest of the cycle. However, surprisingly enough, they play a crucial role, again, during the dark energy-dominated phase at late times.  
As discussed in the next section, IFSs become dynamically important when
their effective friction on the dilaton exceeds the conventional  Hubble friction. 
Immediately after the bounce,  the universe is hot and the Hubble friction dominates over the IFS-induced friction.  Consequently,  the role of the IFSs is negligible for a long period after the bounce. However,  as the universe continues to expand, the Hubble friction decreases, and eventually, at
$H^2 \sim -\frac{ \kappa}{g_s^2} V',$
the IFS-induced friction becomes non-negligible.  

At this stage, something quite dramatic happens. Although the nominal  
dilaton potential $V(\phi)$ is negative, the IFSs induce an additional positive contribution that causes the overall effective  potential to become positive. We refer to this as the {\it IFS-induced dark energy density} which, in the cyclic scenario, accounts for the present phase of accelerated expansion.  In Sections \ref{Instant folded strings} and \ref{LambdaCDM}, we show that the IFS-induced dark energy density has the distinctive feature that it increases for part of the accelerated expansion period before decreasing.  
In the cyclic scenario considered in this paper, the IFS-induced dark energy is responsible for the current acceleration of the universe.

As the universe continues to evolve with $\dot{\phi}>0$, the dilaton is pushed slowly towards strong coupling. Eventually, since the IFS friction scales like $1/g_s$, the Hubble friction becomes the dominant friction again. At this stage, the IFSs once again cease to play an important role, and the dark energy is determined solely by $V<0$.  This naturally causes accelerated expansion to transition to \ch{slow} contraction, 
\ch{a phase extremely efficient in smoothing and flattening the universe. Contraction is considered ``slow" when 
$\varepsilon = -H^2\dot{H}$ is larger than $3$, the value corresponding to kination.}  
The steeper the slope, the slower the contraction and the more effective the smoothing of anisotropies and inhomogeneities. In section \ref{slow contraction}, we discuss some aspects of this stage. In particular, we show that for our potential,  the value of $\varepsilon$ is ${\cal O}(10)$ or greater during the slow contraction phase.

The slow contraction phase persists as long as the negative potential continues to decrease exponentially  at a sufficient rate. Once the field approaches the minimum of the potential, contraction continues but the potential becomes negligible compared to the dilaton kinetic energy density.  A period dominated by dilaton kination ensues. 

As the dilaton evolves past the minimum and up the potential, $V(\phi)$ eventually becomes positive.  Now, a competition emerges between the Hubble anti-friction, which accelerates the dilaton, and the gradient of the potential, which acts to slow it down. At the end of Section \ref{slow contraction} and in the Appendix, we show that, for a sufficiently large potential gradient, the potential ultimately dominates and brings the dilaton to a halt at SP, and a new cycle begins.

\section{Instant folded strings}
\label{Instant folded strings}

IFSs are an essential ingredient in our scenario and especially noteworthy because they are uniquely and explicitly generic features that emerge in string theory when the string coupling grows with time. Unlike most string-inspired elements introduced into cosmology, they cannot be mimicked by conventional (3+1)-dimensional quantum field theory.
In this section, we briefly review their non-standard properties and introduce some of their possible novel effects on cosmology.  More details can be found in \cite{Itzhaki:2018glf,Attali:2018goq,  Itzhaki:2021scf,InstantCosmology}.

\begin{figure}
\includegraphics[width=10cm]{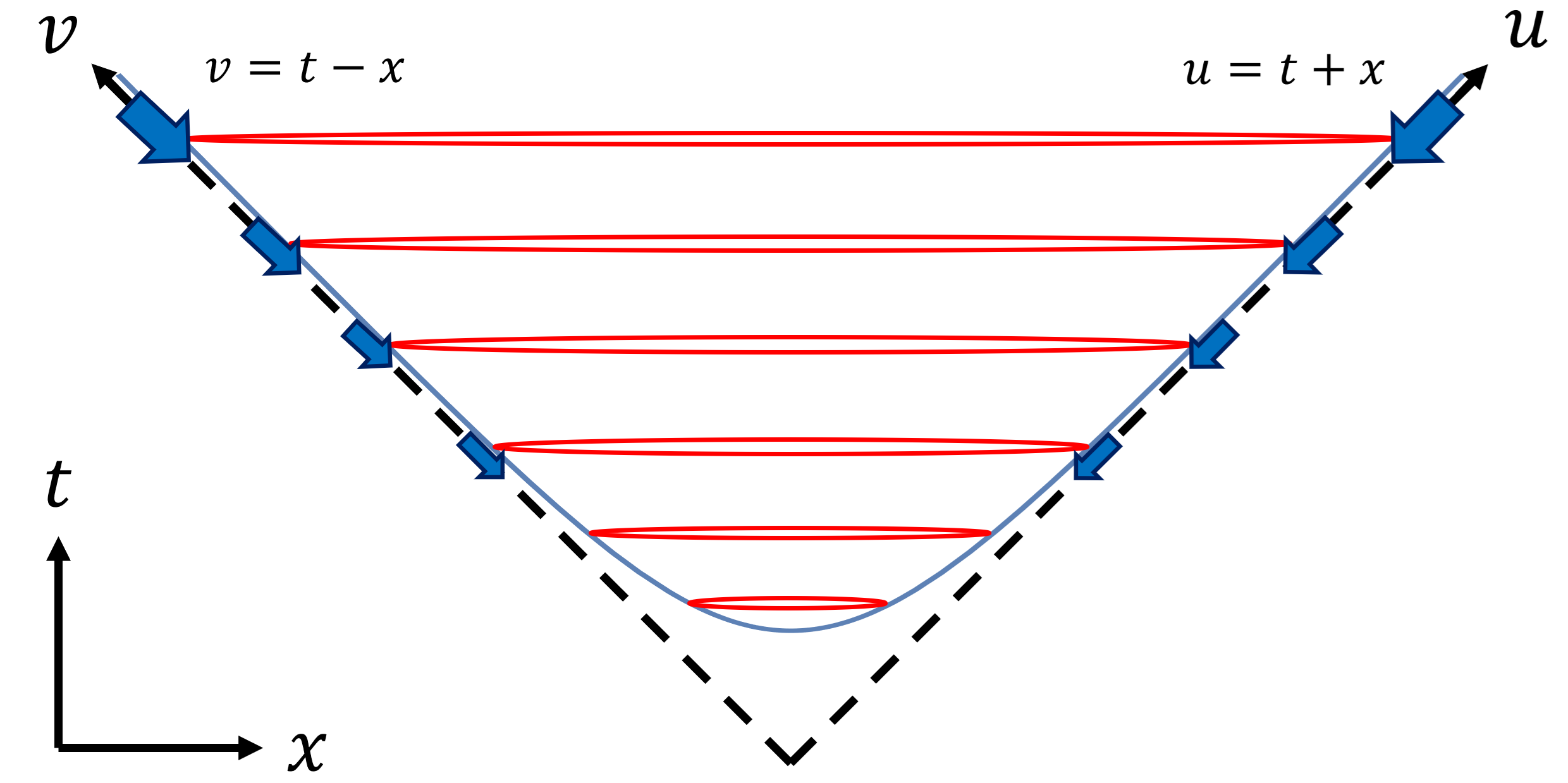}
\centering
\caption{A representation of the solution in which an IFS with a single fold spontaneously appears at time $t_0$ and the fold, at either end, follows a space-like trajectory that asymptotically approaches the speed of light.  
The blue arrows represent the {\it negative} null energy-momentum at the fold that becomes more and more negative over time.
}
\label{fig:IFS}
\end{figure}

IFSs are fundamental closed folded strings with nonstandard features that appear {\em only} when the string coupling grows with time, or, equivalently, when the gradient of the dilaton, $\partial_{\mu}\phi$, is time-like and points to the future. To see what makes the dilaton gradient special, we recall that fundamental strings have to satisfy the Virasoro constraint. To leading order (in the string tension expansion), the Virasoro constraint reads
\be\label{vcon}
g_{\mu\nu}\partial_{\pm}x^{\mu}\partial_{\pm}x^{\nu}=0,
\ee
where $\partial_{\pm}$ are the null derivatives on the string world sheet. Most stringy corrections keep the structure of (\ref{vcon}) intact, leading to modifications that are small when the space-time curvature is small in string units. The dilaton gradient is different in this regard; it adds to  (\ref{vcon}) the term 
\be
\alpha' \partial_{\mu} \phi \partial_{\pm}^2 x^{\mu},
\label{sublead}
\ee
where $\alpha' \equiv \ell_s^2/2\pi$ and $\ell_s$ is the string length. The factor of $\alpha'$ means that (\ref{sublead}) is formally subleading compared to (\ref{vcon}). However, because (\ref{sublead}) is linear in $x^{\mu}$ while (\ref{vcon}) is quadratic,
it can transiently dominate the dynamics governed by (\ref{vcon}), in a region where $\partial_\pm x^{\mu}$ changes rapidly. This brief dominance is sufficient to allow for novel stringy modes that differ significantly from standard closed string excitations and are simply absent when the $\partial_{\mu}$ is not time-like and positive. 
These are the IFSs, which take the unusual form
\be\label{solution}
t(\tau, \sigma)= t_0+ \alpha' Q\log\left( \frac12 \cosh\left(\frac{\tau}{\alpha'Q}\right) +\frac12 \cosh\left(\frac{\sigma}{\alpha'Q}\right)\right) , 
\ee
where $Q$ is the local dilaton time gradient, $Q=\partial_t \phi$ and $x=x_0+\sigma,~y=y_0,~z=z_0.$
This solution describes a closed folded string that is nucleated classically at $(t_0, x_0, y_0, z_0)$. The solution involves a single fold at $\tau=0$, which follows a space-like trajectory that asymptotes to a null trajectory (see Fig. \ref{fig:IFS}). 
The role of $Q > 0$ is twofold. It is necessary for the IFSs to exist and, it ``smooths" the folds, with the folding occurring over a length scale of order $\alpha'Q$, rather than at a singular point.

Since IFSs are produced only when the dilaton varies with time, they play no role in string theory for static backgrounds. However,  IFSs can significantly modify the dilaton-driven equations of motion in  \ch{ cosmology}. To assess their impact, it is essential to determine both their production rate and their energy-momentum tensor. 

The production rate of IFSs was calculated in \cite{Hashimoto:2022dro}. For our purposes, it is more natural to express the result in cosmological conventions in which the dilaton is canonically normalized. This involves rescaling $\phi \to \kappa\phi/\sqrt{2}$, so that the string coupling becomes $g_s = e^{\kappa \phi / \sqrt{2}}$ (as opposed to the standard string theory expression $g_s = e^{\phi}$ used in \cite{Hashimoto:2022dro}). Where $\kappa^2 = 8\pi G_N$ is the gravitational coupling. The production rate is 
\be\label{gwe}
\Gamma_{IFS}=\frac{(\partial_{\mu} \phi)^2}{32\pi^6}\Theta(\dot{\phi}),
\ee
in these conventions. The appearance of the Heaviside Theta function, $\Theta(\dot{\phi})$, indicates that IFSs are only produced when $\dot{\phi}>0$. That is, a striking feature of IFSs is that they generate a built-in violation of time-reversal symmetry.

The energy-momentum tensor associated with an IFS is determined by its solution (\ref{solution}) and was shown to violate the NEC in \cite{Attali:2018goq}.  On  length scales much larger than the characteristic scale $\alpha'Q$ of the fold, the energy-momentum tensor takes the form
\be\label{al}
\begin{aligned}
&T_{uu}=-\frac{v-v_0}{2\pi\alpha'}\Theta(v-v_0)\delta(u-u_0)\delta^{(2)}_\perp(y-y_0,z-z_0), \\
&T_{vv}=-\frac{u-u_0}{2\pi\alpha'}\Theta(u-u_0)\delta(v-v_0)\delta^{(2)}_\perp(y-y_0,z-z_0), \\
&T_{uv}=\frac{1}{2\pi\alpha'}\Theta(u-u_0)\Theta(v-v_0)\delta^{(2)}_\perp(y-y_0,z-z_0).
\end{aligned}
\ee
Here we use  spacetime coordinates $(t, x, y, z)$, with light-cone variables defined as $v \equiv t + x$ and $u \equiv t - x$, and with $\delta^{(2)}_\perp(y - y_0, z - z_0) \equiv \delta(y - y_0)\delta(z - z_0)$. $T_{uv}$ describes the energy density in the bulk of the string, which is positive due to the folded string tension. This positive energy is canceled exactly by the negative null energy at the fold, which is described by $T_{uu}$ and $T_{vv}$. The minus signs in $T_{uu}$ and $T_{vv}$ indicate that, when contracted with null vectors along the $u$ and $v$ directions, respectively,  the IFS violates the null energy condition (NEC) and that the violation occurs at the folds. Moreover, the prefactors $(v - v_0)$ in $T_{uu}$ and $(u - u_0)$ in $T_{vv}$ imply that this NEC violation becomes increasingly significant as the IFS grows in size.  These features are represented by the blue arrows in Fig.~\ref{fig:IFS} that are pointed in the negative null direction and growing in size.

When the lifetime of an IFS is much less than a Hubble time,  
\be\label{yg}
\tau_{IFS} |H| \ll 1,
\ee
as will be the case in the cyclic bouncing model presented in this paper, the total energy density $\rho_{IFS}$ of a uniform gas of IFSs is negligible due to the near cancellation of the positive bulk contribution of an IFS, consisting of a pair of nearly parallel strings segments,  with the negative contribution due to the folds at either end. That fact, combined with the calculation above showing that the NEC is violated, means that the IFSs must have negative pressure. In other words, IFSs induce negative pressure at no energy density cost: 
\be
\rho_{IFS}=0,~~~~p_{IFS}<0.
\ee
In \cite{InstantCosmology} it was shown that  
\be p_{IFS} = -\frac{\gamma \dot{\phi}^2}{3g_s^2},\label{IFS-pressure}\ee
where $\gamma$ is a dimensionless number.

An IFS has a finite lifetime because it can split into smaller pieces, as illustrated in Fig.~4.  
The lifetime of an IFS, $\tau_{IFS}$, is set by its initial splitting time,  $\tau_{IFS} \sim \ell_s / g_s$, as discussed in \cite{InstantCosmology}. Combining this expression with the bound on $\tau_{IFS}$ given in (\ref{yg}), implies our approximations are valid when
\be\label{ond}
 \ell_s H \ll g_s\ll 1.
\ee
The resulting dilaton-gravity equations of motion that describe the cyclic evolution are   
\cite{InstantCosmology}
\begin{subequations}
\label{IFS equations}
\begin{gather}
3H^2 = \kappa^2 \rho_{tot}  \equiv \kappa^2 \left(\rho_r+\rho_m+\rho_{r-IFS} + \frac{1}{2}\dot{\phi}^2 + V(\phi)\right) , \label{a} \\
 6\frac{\ddot{a}}{a} = -\kappa^2(\rho_{tot}+3p_{tot})= \kappa^2 \left[2V + \frac{\gamma\dot{\phi}^2}{g_s^2}\Theta(\dot{\phi}) -2\dot{\phi}^2 - \rho_m - 2\rho_r - 2\rho_{r-IFS}\right], \label{aprime} \\
\ddot{\phi}+3H\dot{\phi} + V'(\phi) =  -\frac{\kappa\gamma}{\sqrt{2}}   \frac{\dot{\phi}^2}{g_s^{2}} \Theta(\dot{\phi}),\label{b} \\ \dot{\rho}_{r-IFS} + 4H\rho_{r-IFS} = \gamma \frac{\dot{\phi}^2}{g_s^{2}}\left(H+\frac{\kappa}{\sqrt{2}}\dot{\phi}  \right) \Theta(\dot{\phi}),\label{c} 
\end{gather}\label{abc}\end{subequations}
where here and throughout the paper dot represents the derivative with respect to time and the prime represents the derivatives with respect to $\phi$. $\rho_r$ and $\rho_m$ are the standard radiation and matter energy density and $V(\phi)$ is the dilaton's potential. $\rho_{r-IFS}$ is the energy density of the radiation that results from the decay of IFSs. Eq. (\ref{c}) implies that, unlike $\rho_m$ and $\rho_r$,  $\rho_{r-IFS}$ can be negative. The microscopic origin of this is the negative null energy at the fold of the IFS. As we shall demonstrate, this allows for the universe to bounce from contraction to expansion.

\begin{figure}[t!]
\includegraphics[width=10cm]{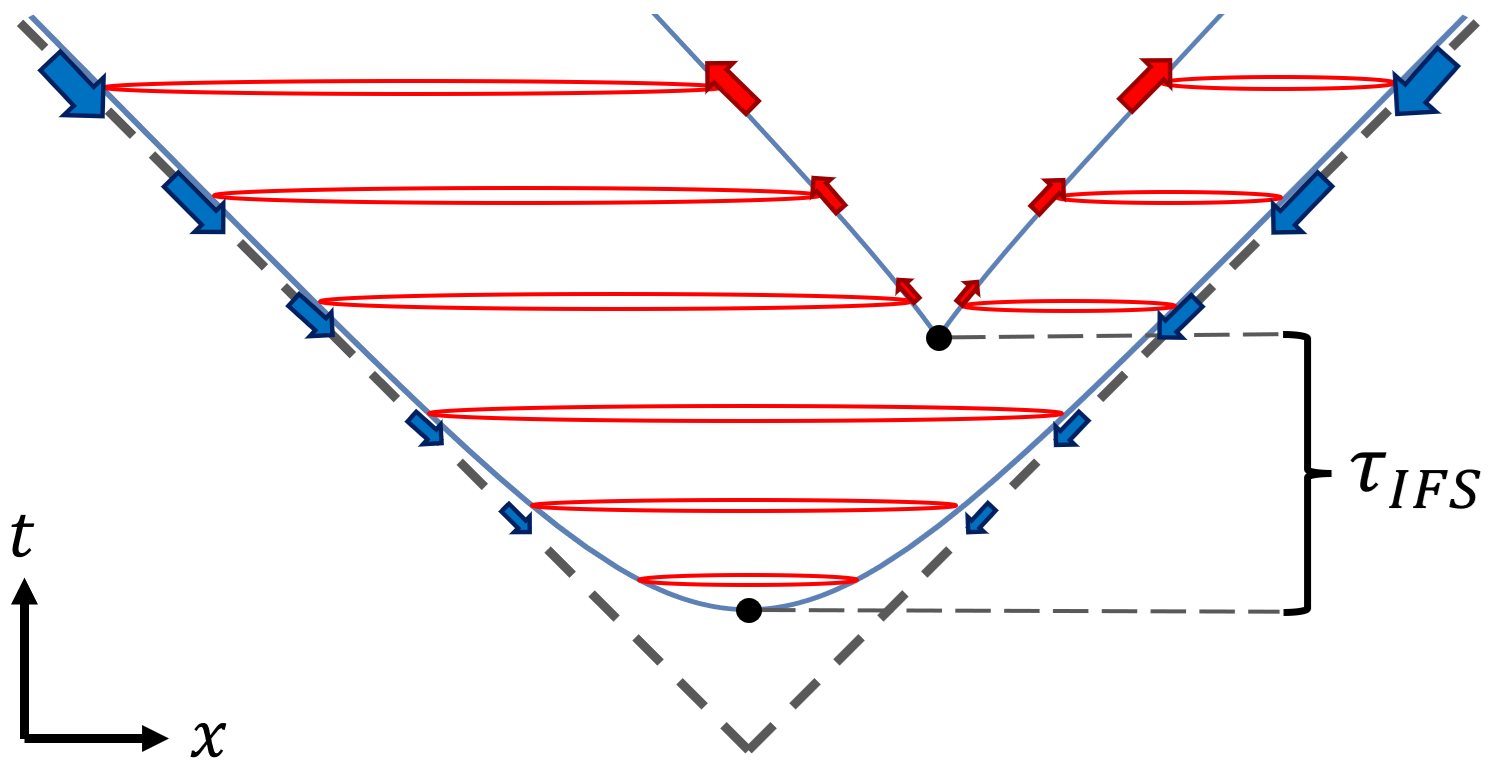}
\centering
\caption{
IFSs decay over time by splitting into smaller segments, as illustrated in the figure.  The original folds have growing negative energy-momentum along the null direction, as indicated by the blue arrows that grow in size along the trajectory, but the new folds generated by the split do not, as indicated by the red arrows. }
\label{fig:IFS split}
\end{figure}

It is instructive to know, without solving  (\ref{IFS equations}), when the IFSs play an important role. In an expanding universe (\ref{b}) provides such a shortcut: we see that when $\dot{\phi}$ and $H$ are positive, the IFSs induce an extra friction term (rhs of (\ref {b})) which ``competes" with the Hubble friction ($3 H \dot{\phi}$).  
When the Hubble friction dominates, the IFSs can be ignored.  In contrast, when the IFS friction dominates, something remarkable happens: an IFS-induced dark energy-dominated phase, one that could potentially explain the period of accelerated expansion observed at present, as detailed in section 6.

During an IFS-induced dark energy-dominated phase,  (\ref{b})  reduces to  $V'(\phi) \approx  -\frac{\kappa\gamma}{\sqrt{2}}   \frac{\dot{\phi}^2}{g_s^{2}}$ and (\ref{c}) reduces to $\rho_{r-IFS} = \gamma \frac{\dot{\phi}^2}{4 g_s^{2}}$.  As a result, the first Friedmann equation (\ref{a}) can be approximated as 
\be
3H^2 \approx \kappa^2\left(V -\frac{1}{\sqrt{8}\kappa}V'\right). \label{Hsqr}
\ee
In this limit, the effective  dark energy density is 
$V -\frac{1}{\sqrt{8}\kappa}V'$, which, unlike conventional quintessence models, includes a contribution that depends on $V'$ as well as $V$. 

The dependence on $V'$ 
introduces two novel effects that play important roles in the dark-energy dominated phase of our cyclic model.  First, even if the dilaton potential is negative, $V<0$, the effective IFS-induced dark energy can be positive for some choices of $V$. A simple example is  $V(\phi) = - V_0 \exp\left(\lambda \kappa \phi\right)$ where $V_0 >0$ and $\lambda > \sqrt{8}$.  Secondly, 
as this same example illustrates, if $\dot{\phi}>0$, the dark energy density can increase with time.

\section{Crossing the ESP}
\label{crossing}
As the dilaton moves towards weak coupling ($\dot{\phi}<0$) during the contracting phase, its kinetic energy is being blue-shifted, which drives the universe towards a singularity which must be avoided; and, because the kinetic energy grows so rapidly, a device designed to bring the dilaton to a halt and cause it to reverse direction must act very rapidly.  In this section, we describe in detail how a dilaton passing through an Enhanced Symmetry Point (ESP) in a contracting universe is a natural mechanism for achieving these necessary conditions.

As described in \cite{ESP}, the number density of $\chi$ particles being produced, $n_{\chi}$, is related to the velocity at which the dilaton crosses the ESP
\be n_\chi^{(ESP)} = \frac{(\lambda_{\chi}\dot{\phi}_{ESP})^{3/2}}{(2\pi)^3}.\ee
Their coupling with the dilaton (\ref{f}) implies that the $\chi$ particles induce a time-dependent potential (on top of the time-independent potential $V(\phi)$) 
\be V_{\chi}(\phi) = \lambda_{\chi} n_\chi(t) |\phi-\phi_{ESP}|, \quad n_\chi(t) = \frac{n_\chi^{(ESP)}}{a^3(t)}, \label{chi potential}\ee 
where the scale factor is normalized such that $a(t)=1$ at the time of ESP crossing.

Since $V(\phi)$ is negligible compared to the kinetic energy of the dilaton over the range of $\phi$ near the ESP, the dilaton equation of motion simplifies to
\be\label{li} \ddot{\phi} + 3H\dot{\phi} + V'_{\chi}(\phi) = 0,\ee
which implies that 
\be \frac{d}{dt}\left( a^3\dot{\phi} \right)=-\lambda_{\chi} n_\chi^{(ESP)} = -\frac{1}{(2\pi)^3}\lambda_{\chi}^{5/2} \dot{\phi}_{ESP}^{3/2},
\label{re1}\ee
which is a constant.
This is a result of $\dot{\phi}$ and $V_\chi$ both scaling together as $1/a^3(t)$.

This implies that the time it takes for the dilaton to stop decreasing, $t_*$, is independent of the Hubble parameter $H$ 
\be t_* \equiv (2\pi)^3 \lambda_{\chi}^{-5/2} \dot{\phi}_{ESP}^{-1/2}, \label{stopping-time}\ee 
and that 
\be \dot{\phi}(t) = \left(1-\frac{t}{t_*}\right)\frac{\dot{\phi}_{ESP}}{a^3(t)}, \label{stopping phi dot}\ee
which indeed vanishes at the `stopping time' $t_*$.

\begin{figure}[t!]
\includegraphics[width=14cm]{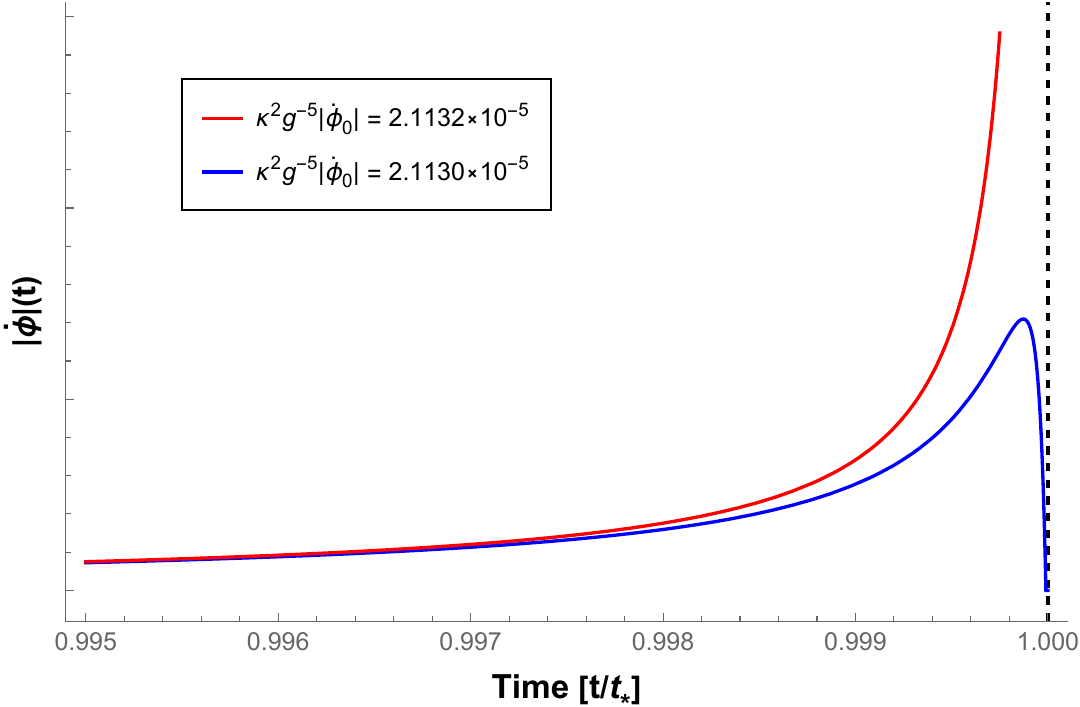}
\centering
\caption{Criticality test. The different behavior of $\dot{\phi}(t)$, just above, and just below, the critical value of $|\dot{\phi}_{ESP}|$.
The x-axis shows time, rescaled as $t/t_*$, while on the y-axis we have the magnitude of $\dot{\phi}$.}
\label{fig:critical}
\end{figure}

Equipped with (\ref{stopping-time}), we are in a position to address the concerns raised in Section \ref{nutshell}. 
The time it takes for a contracting universe to collapse to a singularity is of order $1/H_{ESP}$. Thus, the rough bound on $\dot{\phi}_{ESP}$, needed to ensure the dilaton stops prior to the collapse, is
\be H_{ESP}\, t_* \approx \frac{1}{\sqrt{6}}\kappa\,\dot{\phi}_{ESP} \, t_* \lesssim 1; \ee
a numerically precise version of this estimate is
\be\label{fc} \dot{\phi}_{ESP} \leq 2.1131 \times 10^{-5} \lambda_{\chi}^5 \, \kappa^{-2}.\ee
Fig.~\ref{fig:critical} illustrates the different behavior just above and just below the critical value of $\dot{\phi}_{ESP}$. Above criticality, $\dot{\phi}$ diverges as the universe collapses before reaching $t_*$. For the initial value chosen just shy of criticality, $\dot{\phi}$ grows substantially, but manages to flip its sign and halt just before the collapse.

The energy density of the dilaton is roughly the same as the energy density of the $\chi$ particles that reheat the universe. Consequently, (\ref{fc})  implies that 
\be
T_{rh} \lesssim \lambda_{\chi}^{5/2} 10^{16} \mbox{GeV} < 10^{16} \mbox{GeV}, \label{ESP bound}
\ee
is a rough upper bound on the reheating temperature.

\section{The bounce}
\label{bounce}
In Section \ref{nutshell}, we presented an argument based on the Bianchi identity explaining why IFSs can trigger a bounce so efficiently. The key insight lies in the fact that the NEC violation induced by the IFS is amplified during a contracting phase. This feature — embodied by the dependence of the NEC-violating component on the scale factor $a(t)$ —  is crucial for the analysis below. As we shall see, in a contracting universe, the dominant NEC-violating component grows more rapidly than all other sources, which enables the bounce.
Like in \cite{InstantCosmology}, the bounce described here is smooth and non-singular.  
However, there are some key differences from the toy model considered in \cite{InstantCosmology}. One important difference is that IFSs in the toy model were always present, whereas here IFS production begins only when $\dot{\phi}$ switches from negative to positive. 
This switch initiates the transition from a standard contracting phase with no NEC violation, where both $H$ and $\dot{H}$ are negative, through a sequence of NEC violating phases  
in which $\dot{H}$ changes sign, then $H$ does, and then $\dot{H}$ changes sign once again.
Another significant difference is that, in the cyclic scenario, the potential driving the dilaton, $V_\chi$, is time-dependent. 

As explained in the previous section, the kinetic energy density of the dilaton vanishes and the standard time-independent potential $V(\phi)$ is negligible at the stopping time, meaning that the \ch{driving} potential for $\phi$  is now entirely dominated by the positive energy density associated with the $\chi$ particles 
\be \rho_{\chi} =  V_\chi(\phi_*) = \frac{ \lambda_{\chi} n_\chi^{(ESP)}}{a^3(t_*)}|\phi_*-\phi_{ESP}|. \ee
The universe is contracting at this stage, which means that this positive energy density will continue to grow with time like $1/a^3(t)$.
Since the slope of the effective potential is negative,
\be\label{hi}
 V_\chi' =- \frac{ \lambda_{\chi} n_\chi^{(ESP)}}{a^3(t)},
\ee 
$\dot{\phi}$ becomes positive, which triggers the production of IFSs. 
The IFS friction keeps the dilaton's speed small, only allowing it to grow until an equilibrium attractor solution is reached in which the gradient of the potential pushing the dilaton forward balances the IFS friction holding it back. 

This limit was thoroughly explored in \cite{InstantCosmology}, albeit in a slightly different context with a standard time-independent dilaton potential, and was termed the IFS slow-roll limit.
In the IFS slow-roll limit, the dilaton equation and the radiation continuity equation can be approximated by
\begin{subequations}
\begin{align}
V_{\chi}'(\phi) &\approx  -\frac{\kappa\gamma}{\sqrt{2}}   \frac{\dot{\phi}^2}{g_s^2}, \label{slow-roll phi}\\
\dot{\rho}_{r-IFS} + 4H&\rho_{r-IFS}   \approx - \frac{\sqrt{2}}{\kappa} H V'_{\chi}(\phi),\label{slow-roll rad}
\end{align}
\end{subequations}
to leading order in the $\gamma^{-1}g_s^2 \ll 1$ limit. 
The time it takes to reach the slow-roll attractor solution is much shorter than the Hubble time, of the order of
\be t_{SR} \sim \frac{g_s}{\kappa \sqrt{\gamma}}(V'_\chi)^{-\frac{1}{2}} \sim \frac{g_s}{\sqrt{\gamma}}\left(\kappa|\phi_*-\phi_{ESP}|\right)^{\frac{1}{2}} H^{-1}, \ee
implying that we can treat this transition as effectively instantaneous.

\begin{figure}[!t]
\includegraphics[width=13cm]{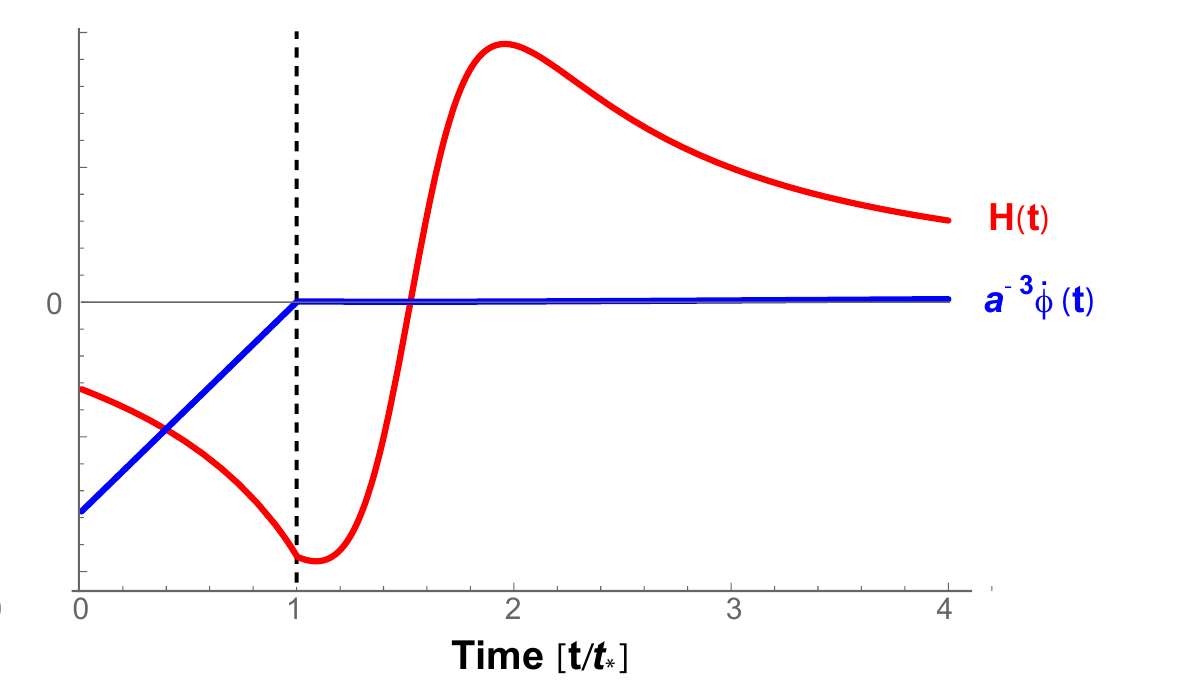}
\centering
\caption{Halting the dilaton and the ensuing bounce. On the x-axis we have time, beginning from the moment of  ESP crossing and measured as a fraction of the stopping time $t_*$. The y-axis depicts the Hubble rate, $H$, in red as well as the speed of the dilaton, rescaled as $a^3|\dot{\phi}|$, in blue. The momentum of the field, $a^3|\dot{\phi}|$, decelerates at a constant rate according to (\ref{re1}), until stopping time. After the stopping time, IFS friction keeps the speed and momentum of the dilaton negligible, but greater than zero. The NEC violation induced by the IFS and their decay products can flip the sign of $\dot{H}$ and then eventually $H$, leading to a bounce. In the particular example shown, $\gamma^{-1}g_s^2 \sim 10^{-5}$, thus the slow-roll approximation is valid.}
\label{fig:ESP and bounce}
\end{figure}

Note that, given $V_\chi' < 0$, the source term on the right-hand side of (\ref{slow-roll rad}) is negative in a contracting universe. This reflects the fact that the decay products of the IFS carry negative energy in such a background, thereby contributing to the violation of the NEC. Since these decay products are radiation-like, their energy density scales as $1/a^4$ during contraction, making them the dominant source of NEC violation. During slow-roll, the kinetic energy density of the dilaton, $\frac{1}{2} \dot{\phi}^2$, remains negligible compared to $V_\chi$ because IFS friction slows the dilaton down. As the universe continues to shrink, the NEC-violating component $\propto 1/a^4$ will always overtake the potential $V_\chi \propto 1/a^3$, ultimately triggering a bounce.

We can make this argument more precise by solving equation (\ref{slow-roll rad}) explicitly, yielding
\be
\rho_{r-IFS}(t) \approx  \frac{\sqrt{2}}{\kappa}n_\chi^{(ESP)}\left(\frac{1}{a^3(t)}-\frac{a_*}{a^4(t)} \right),
\ee
The stopping-time $t_*$, which is roughly when slow-roll begins, serves as an initial condition at which  $\rho_{r-IFS}(t_*)=0$.
We can see that in a contracting universe, the energy density of the decay products indeed becomes negative, and that the dominant term scales like $ 1/a^4$.

The total energy density, which is a combination of $V_\chi$ and $\rho_r$, with negligible contribution from the dilaton kinetic term and the IFSs themselves, determines the rate of contraction,
\be 3H^2 \approx \kappa^2(\rho_\chi+\rho_{r-IFS}) \approx  \sqrt{2}\kappa n_\chi^{(ESP)}\left[ \left(1 + \frac{g\kappa|\phi_*-\phi_{ESP}|}{\sqrt{2}}\right)\frac{1}{a^3(t)}-\frac{a_*}{a^4(t)} \right], \label{nice bounce}\ee
we see that contraction must stop, at some time $t_b$, when the scale factor has contracted to the size of
\be a_b \equiv a(t_b) \approx a_*\left(1 + \frac{g\kappa|\phi_*-\phi_{ESP}|}{\sqrt{2}}\right)^{-1}.\ee
At this point, we have $H=0$, while $\dot{H}>0$, meaning that $t_b$ is the moment of the bounce, when the universe transitions from contraction to expansion.
Notably, the bounce in (\ref{nice bounce}) is non-singular, with finite curvature at $t_b$ 
\be R(t_b)\equiv [6\dot{H}+12H^2]_{t_b} = \frac{\sqrt{2}\kappa a_*}{a_b^{4}}n_\chi^{(ESP)}.\ee
Fig.~\ref{fig:ESP and bounce} shows an exact numerical solution to the equations in (\ref{abc}), given the $\chi$ induced potential (\ref{chi potential}), which is in agreement with the slow-roll approximation.

\section{An IFS-induced dark energy phase}
\label{LambdaCDM}

As discussed in Section \ref{nutshell},
shortly after the bounce, the Hubble friction dominates the IFS-induced friction, and the IFSs play no significant role in the evolution of the universe until the dark energy phase approaches.
The universe enters a conventional radiation-dominated phase filled with hot standard matter particles and radiation that is followed by a conventional matter-dominated phase. During these two phases, since $V<0$ and the IFSs play no role, the dark energy density is negative but tiny, so small compared to the matter and radiation densities that it has no observational consequences. 

This situation remains until redshift 
\be
z_{IFS}\sim  \gamma^{1/3}g_s^{-2/3}, 
\label{zifs}
\ee
at which time, according to (\ref{c}), the Hubble friction has decreased sufficiently and $\dot{\phi}>0$ has increased sufficiently
that IFSs production is no longer suppressed and the associated IFS-friction becomes non-negligible. To be consistent with cosmological observations, $z_{IFS}$ must be greater than 1, which requires that  $g_s^2\ll \gamma.$

At this stage, the IFSs induce a contribution to the effective dark energy, as described at the end of Section~\ref{Instant folded strings}.
To analyze this contribution, let us examine the trace of the energy-momentum tensor associated with the dark energy sector, which can be expressed as 
\be -\kappa^2 (T^{\mu}_{~\mu})_{DE} \equiv 6\dot{H} + 12H^2 - \kappa^2\rho_m =4\kappa^2\left[V-\frac{1}{\sqrt{8}\kappa}V' - \frac{1}{4}\dot{\phi}^2 - \frac{1}{\sqrt{8}\kappa}\ddot{\phi}\right] - 3\sqrt{2}\kappa H\dot{\phi}. 
\label{trace eq}\ee
The first equality follows from the relation between the Ricci scalar, $R \equiv 6\dot{H}+12H^2$, and the trace of the energy-momentum tensor. Since radiation is traceless, what remains once matter has been separated out is the trace contribution of the dark energy sector, denoted here by $(T^{\mu}_{~\mu})_{DE}$.
The second equality is obtained by taking a particular linear combination of equations (\ref{a}) through (\ref{b}) that gives $6\dot{H}+12H^2 -\kappa^2 \rho_m$.

Eq.~(\ref{trace eq}) is an 
exact relation that is valid in dilaton gravity with or without 
IFSs. In the absence of IFSs, the terms proportional to $V'$, $\ddot{\phi}$, and $H \dot{\phi}$ sum to zero according to the dilaton equation, so (\ref{trace eq})
reduces to the familiar form of the trace equation $6\dot{H}+ 12H^2=\kappa^2(4V+\rho_m-\dot{\phi}^2)$.
In contrast, during periods when the IFSs \ch{and the friction they induce on the dilaton} dominate, the conditions change such that   \be 
\sqrt{2}\kappa^{-1} V'(\phi) \approx  -\frac{\gamma}{g_s^{2}} \dot{\phi}^2, \;\; 
\ee
\ch{which implies that, for $g_s^2\ll \gamma$, the $\dot{\phi}^2$, $\ddot{\phi}$ and $H \dot{\phi}$ terms are negligible}
and (\ref{trace eq}) takes on a particularly simple form, 
\be 6\dot{H}+12H^2 = \kappa^2\rho_m + 4\kappa^2 V_{\text{eff}} \quad {\rm where} \quad V_{\text{eff}} \approx  V-\frac{1}{\sqrt{8}\kappa}V', \label{veff}\ee 
and we see that, as discussed in section 3, $V'$ contributes to the IFS-induced dark energy in such a way that $V_{\text{eff}}$ can become positive even though $V$ is negative.

It is not difficult to find negative potentials whose slope is sufficiently steep that $V_{eff}$ is positive.  However, we have decided for the example in this paper to restrict consideration to  potentials that can be expressed as a power-law expansion in $g_s =e^{\frac{\kappa\phi}{\sqrt{2}}}$, {\it i.e.}, 
\be
V(\phi)=\sum_{j=1}c_j g_s^{2+2j} = \sum_{j=1} c_j  e^{\frac{(2+2j)\kappa\phi}{\sqrt{2}}}
\label{vgexp}
\ee
since potentials of this form are well motivated in string theory, as they naturally arise in the perturbative expansion and do not require any non-perturbative contributions.
In this particular case,  the leading order ($j=1$) term $\propto g_s^4$  contributes precisely zero to $V-\frac{1}{\sqrt{8}\kappa }V'$ in (\ref{veff}).     
As a result, conditions $\dot{\phi}^2,\ddot{\phi}\ll \kappa^{-2}H^2$  
become nontrivial, since the terms on the left and right sides are now of the same order $g_s^6$. Hence, to simplify the analytical discussion, we restrict ourselves to potentials where
\be c_1,c_2<0 ~~~\text{and}~~~ \mu \equiv \frac{\gamma c_2}{c_1} \gg 1,\ee
for which $\frac{\kappa^2\dot{\phi}^2}{3H^2} \sim \frac{8}{\mu}\ll1$ and $\frac{\kappa^2\ddot{\phi}}{3H^2} \sim \frac{3\sqrt{8}}{\mu}$ so that 
(\ref{veff}) remain valid in the weak coupling limit, and to leading order (both in $\mu^{-1/2}$ and $\gamma^{-1}g_s^2$), one obtains the IFS-induced dark-energy density
\be \rho_{DE} = \frac{1}{2}c_2 g_s^6 + \mathcal{O}(\mu^{-1/2}, g_s^8). \label{DE density}\ee
The dark energy density in (\ref{DE density}) grows 
as $g_s$ increases, leading to the following correction to $w_{DE}$ \cite{InstantCosmology} 
\be w_{DE}(\mu, g_s) = -1 - \sqrt{48\mu^{-1} + 288\gamma^{-1}g_s^2} + \mathcal{O}(\mu^{-1},\gamma^{-1}g_s^2)< -1. \label{w g mu}\ee
That is, the IFS-induced dark-energy density is generally slowly increasing, but acts like a cosmological constant, with $w_{DE} \to -1$, at the limit where $\mu \gg 1$ and $g_s^2 \ll \gamma$.
Eq.\,(\ref{w g mu}), together with (\ref{DE density}), also implies that
\be \frac{d \ln(g_s)}{d \ln(a)} = \frac{1}{3}\frac{d \ln(H)}{d \ln(a)} = \frac{1}{2}(1-w_{DE}) = \sqrt{12\mu^{-1} + 72\gamma^{-1}g_s^2} + \mathcal{O}(\mu^{-1},\gamma^{-1}g_s^2). \label{g_s evolution}\ee
Since $\mu\gg 1$ and $g_s^2 \ll \gamma$, the value of $w_{DE}(\mu, g_s)$ also changes very slowly, so that for some number of $e$-foldings in $a$, $w_{DE}$ will remain nearly constant.

\begin{figure}[!t]
\includegraphics[width=12cm]{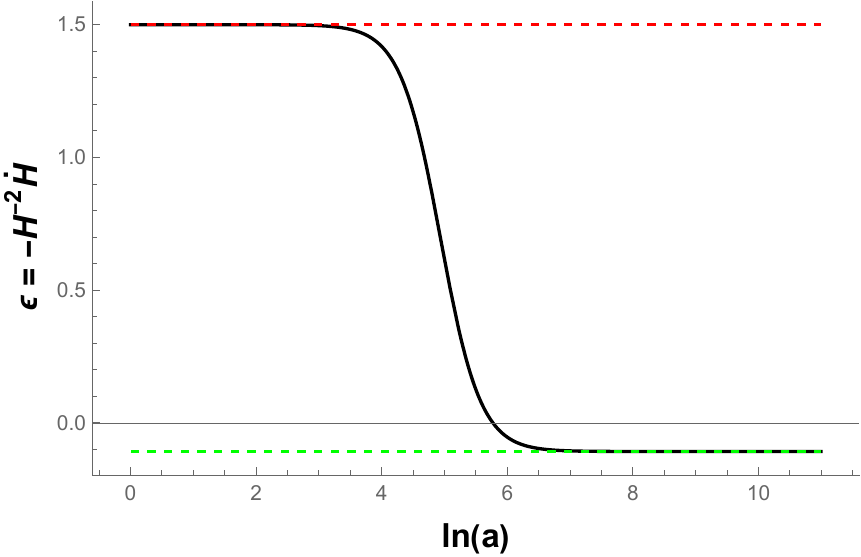}
\centering
\caption{The parameter $\epsilon =\frac{3}{2}(1+w)$ shown as a function of the number of $a(t)$ $e$ foldings, starting in a matter-dominated universe, and ending in a phase of IFS-induced dark-energy domination. Shown here for the values $\gamma^{-1} g_s^2 \sim 10^{-8}$ and $\mu = 10^4$.
}
\label{fig:mdet}
\end{figure}

Going beyond the analytical discussion, we can solve the exact equations numerically.
In Fig.~\ref{fig:mdet} we plot the parameter $\epsilon \equiv -H^{-2}\dot{H} \equiv \frac{3}{2}(1+w)$, as it evolves from $\epsilon=3/2$, during matter domination, to a negative value given approximately by $\frac{3}{2}(1+w_{DE})$, as indicated by (\ref{w g mu}), during IFS-induced dark energy domination.
As $w_{DE}\to -1$, Fig.\,\ref{fig:mdet} exactly replicates the matter-dark-energy transition in $\Lambda$CDM. In particular, when
\be \mu \gtrsim 10^3 \; \; {\rm and} \; \; \gamma^{-1} g_s^2 \lesssim 5\times 10^{-4}, \label{rough limits}\ee
IFS-induced dark-energy in our model becomes indistinguishable from a cosmological constant (with current cosmological observations).

\begin{figure}[t!]
\includegraphics[width=11cm]{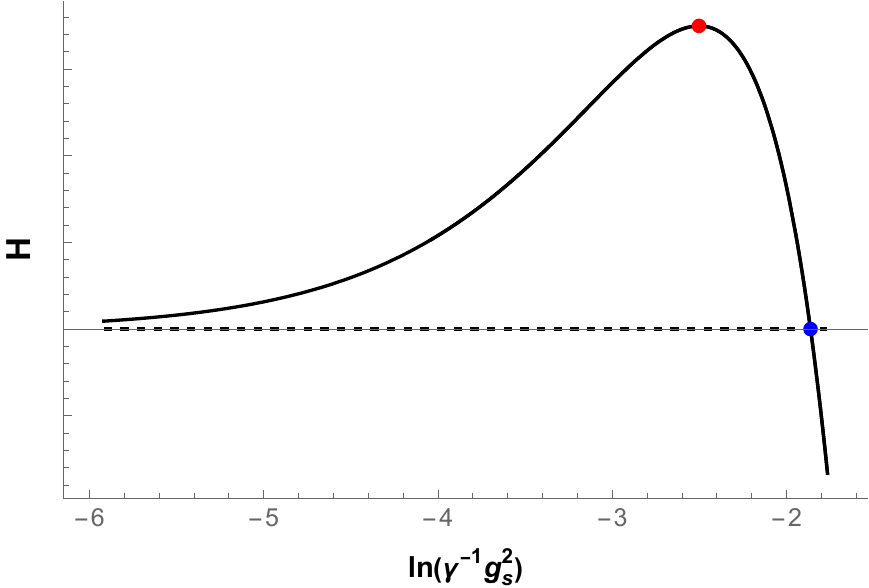}
\centering
\caption[Caption for ETC]{A plot of the numerical solution to $H$ as a function of $\log(\gamma^{-1} g_s^2)$, where $\mu \gg 1$. The Hubble parameter reaches its maximum value (red circle), $H_{\text{max}} \approx1.75 \times 10^{-3} ~ c_2^{1/2} \gamma^{3/2}$, at $\gamma^{-1} g_s^2 \approx 8.18 \times 10^{-2}$, and then begins to decrease, eventually transitioning from positive to negative at $\gamma^{-1} g_s^2 \approx 0.155$ (blue circle).
}
\label{fig:ETC}
\end{figure}

In Fig.~\ref{fig:ETC}, we
plot the numerical value of $H$ as a function of the parameter $\gamma^{-1} g_s^2$. Since $\gamma^{-1} g_s^2$ is monotonic with time, this figure also captures the time-evolution of $H$.
In particular, since the dilaton is slowly rolling towards strong coupling, eventually $g_s^2\sim \gamma$, and the leading-order analysis is invalid. As IFS-induced friction no longer dominates, the IFSs contribution to the dark energy subsides, and since $V(\phi$) is negative, the universe eventually enters a period of contraction (described in the next section).

This raises the question: how many $e$-foldings does the IFS-induced dark energy last? 
A rough estimate of the number of $e$-foldings of $a(t)$ can be obtained from (\ref{g_s evolution})
\be \# e\text{-foldings of } a(t) \lesssim 
\frac{\gamma^{1/2}g_s^{-1}|_{\text{MDE}}}{6\sqrt{2}}, 
\label{a estimate}\ee
where MDE denotes the time of the matter-dark energy transition, the start of the IFS-induced dark-energy-dominated phase.
During this phase, $H(t)$ is also growing, but more slowly,
\be \#  e\text{-foldings of } H(t) \approx -\frac{3}{2}\ln(\gamma^{-1}g_s^2|_{\text{MDE}})-6, \label{H estimate}\ee
where we have used the leading behavior of $H^2 \approx \frac{1}{2} c_1 g_s^6$ at the $\gamma^{-1}g_s^2 \ll 1$ limit, as well the estimate $H_{\text{max}} \approx1.75 \times 10^{-3} ~ c_2^{1/2} \gamma^{3/2}$, for the maximum value of $H$ obtained numerically.

\section{The slow contraction phase}
\label{slow contraction}

Slow contraction begins at the end of the IFS-induced dark energy–dominated phase described in Section~\ref{LambdaCDM}. The dark energy-dominated phase is driven by the influence of IFSs and their decay products on the equation of state, which raises the effective potential $V_{eff}$ to the positive value given in (\ref{DE density}). The expansion continues until $g_s^2 \sim \gamma$, at which point IFS production is suppressed and Hubble friction dominates over IFS friction once again. The dilaton is then free to roll down the steep negative potential and initiate the slow contraction phase.

Assuming $\gamma$ is small, the transition to the slow contraction phase occurs at weak coupling, and the slow contraction phase continues as the dilaton traverses a substantial range of $\phi$ that is all in the weak coupling regime  ($g_s \ll 1$) and, hence, under perturbative control throughout. In this regime, we expect the potential to take the perturbative form $$V = -c_1 g_s^4-c_2g_s^6+\mathcal{O}(g_s^8),$$ as discussed in Section~\ref{LambdaCDM}. Recall that consistency with cosmological observations requires the parameter $\mu \equiv \gamma c_2 / c_1$ to be large.  This choice ensures that, by the time $g_s^2 \sim \gamma$, the two-loop term $-c_2 g_s^6$ already dominates the one-loop contribution $-c_1 g_s^4$. Consequently, the potential in the slow contraction regime can be approximated by
\be V(\phi) \sim -c_2 g_s^6 = -c_2 e^{3\sqrt{2} \kappa \phi}, \label{2loop potential} \ee
up to higher-order corrections.

\begin{figure}[!t]
\includegraphics[width=12cm]{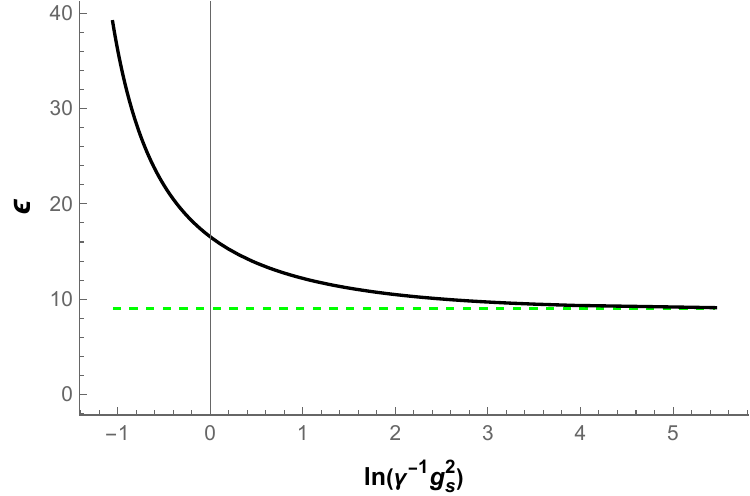}
\centering
\caption{Evolution of the parameter $\varepsilon = -H^{-2} \dot{H}$ as a function of $\ln(\gamma^{-1} g_s^2)$, shown after $H$ becomes negative, in the large $\mu$ limit. The expected asymptotic value $\varepsilon \to 9$ is indicated by the green line and agrees well with the numerical result.}
\label{fig:epsilon} \end{figure}

The equation of state of the attractor solution for the dilaton evolving down the negative exponential potential,  $\varepsilon$, is 
$\varepsilon = \frac{1}{2} \alpha^2$, which is equal to 9 for the potential in (\ref{2loop potential}). 
Figure~\ref{fig:epsilon} confirms this expectation.
If higher order terms in the perturbative potential also come with a negative sign, and eventually come to dominate the two-loop term, it is also possible to obtain a larger value of $\varepsilon$. For example, a potential dominated by the $n^\text{th}$ loop term $V \sim -c_n g_s^{2(1+n)} = -c_n e^{\sqrt{2}(1+n)\kappa \phi}$ yields, respectively,
\be \varepsilon_{n} \cong (1+n)^2.\ee
Based on the analysis of~\cite{Cook:2020oaj}, a value of $\varepsilon \geq \mathcal{O}(10)$ is sufficient for slow contraction to smooth and flatten the universe. 

In examples of the cyclic scenario in the literature, such as \cite{Ijjas:2019pyf}, the maximum (positive) value of $H$, which occurs right after the bounce, is roughly equal to $|H|$ when $H$ is most negative, just before the bounce. Our scenario exhibits the same characteristic behavior, as can be seen in Fig.~\ref{fig:ESP and bounce}. In fact, in the $\gamma g_s^2 \ll 1$ limit, in the region where $\dot{\phi} > 0$ (to the right of the dashed line marking the stopping time) $H$ is nearly antisymmetric around the bounce. This implies that the magnitudes of $H_{\text{max}}$ and $H_{\text{min}}$ are approximately equal.

For example, if the reheating temperature following the bounce is  $10^{8}$GeV, the total decrease in $H$ from that time to the present, when the universe’s temperature is about $10^{-13}$GeV, corresponds to around 100 $e$-folds. The change in $H$ during the IFS-induced dark energy-dominated phase can be relatively small, as shown in Eq.~(\ref{H estimate}). As a result, during the contraction phase, $|H|$ can increase by nearly the same amount, up to 100 $e$-folds.  
Of these, a fraction $\frac{\sqrt{\varepsilon}}{\sqrt{\varepsilon} + \sqrt{3}}$ of the $e$-folds of increasing $|H|$ occurs during slow contraction, when the dilaton rolls down a steep negative potential and $\varepsilon \gtrsim 9$. The remaining fraction, $\frac{\sqrt{3}}{\sqrt{\varepsilon} + \sqrt{3}}$, occurs after the dilaton passes the minimum of the potential. In this stage, the potential energy becomes negligible compared to the blueshifting kinetic energy density of the dilaton, and the universe enters a phase of kination contraction with $\varepsilon \to 3$.
The same considerations also give us an estimate for the depth of the potential at the minimum, namely,
\be \label{V_min}
|V_{min}| \sim \left(|H_0|^{\sqrt{3}} \cdot |H_{max}|^{\sqrt{\varepsilon}}\right)^{\frac{2}{\sqrt{3}+\sqrt{\varepsilon}}} \gtrsim (1\,\text{GeV})^4. \ee
After the bounce, $H$ decreases more rapidly than the scale factor $a$, scaling as $H \sim a^2$ during radiation domination and $H \sim a^{3/2}$ during matter domination. As a result, quantum perturbation modes that exited the Hubble radius during the last $\sim 50-60$ $e$-foldings of contraction of the Hubble radius ($|H|^{-1}$) eventually re-enter it during the radiation- and matter-dominated expanding phases and become responsible for the evolution of large-scale structure and CMB temperature fluctuations. 
This requires that sufficient smoothing must have occurred during the first $\sim 40-50$ $e$-foldings of $H$ to ensure compatibility with present-day observations. 

In this paper, we do not describe in detail how quantum fluctuations of fields during slow contraction are ultimately converted to density perturbations.  There is nothing novel in these aspects compared to mechanisms described in  \cite{Creminelli:2007aq,Levy:2015awa,Brandenberger:2020ekpyrotic,Ijjas:2021ewd}.  In particular, the approach of adding a second scalar field \ch{kinetically} coupled to $\phi$, as described in \cite{Levy:2015awa,Ijjas:2021ewd}, can be directly applied to our scenario without affecting the cosmological background evolution described in this paper. 

Another aspect of the scenario we need to address is arranging a stopping point SP where $\dot{\phi}=0$, a new cycle begins, and the dilaton begins to move back towards weak coupling ($\dot{\phi}<0$). The question is whether the potential that appears in our scenario, can overcome the Hubble anti-friction and stop the dilaton. To see that this is possible, consider the dilaton equation in a contracting universe dominated by the dilaton and its potential:
\be \ddot{\phi} + 3H\dot{\phi} + V' =\ddot{\phi} - \kappa\sqrt{3V+\frac{3}{2}\dot{\phi}^2}~\dot{\phi} + V' = 0. \label{explicit}\ee
If $R$ is the ratio of dilaton kinetic to potential energy,  and $\alpha$ is the slope of the potential,  
\be R \equiv \frac{\dot{\phi}^2}{2V}, \;\; {\rm and} \; \;  \alpha \equiv \frac{V'}{\kappa V},  \label{R alpha def}\ee
then 
\be \kappa^{-1}\frac{dR}{d\phi} \equiv \kappa^{-1}\dot{\phi}^{-1}\frac{dR}{dt}= \frac{\dot{\phi}\ddot{\phi}}{\kappa V} - \frac{V'\dot{\phi}^2}{2\kappa V^2} = 
\sqrt{6R^2 + 12R} - \alpha R - \alpha. \label{R eq}
\ee
According to equation (\ref{R eq}),  $\frac{dR}{d\phi}<0$ if and only if
\be 6R^2+12R < \alpha^2(R+1)^2. \label{stopping condition}\ee
For $\alpha < \sqrt{6}$, (\ref{stopping condition}) is only valid if $R$ is sufficiently small. For $\alpha > \sqrt{6}$, which is more relevant to our perturbative potential,
we obtain from (\ref{R eq}) the inequality:
\be
\kappa^{-1}\frac{dR}{d\phi} \leq (\sqrt{6} - \alpha)(1 + R). \label{R ineq}
\ee
This implies that
\be
-\ln\left(1 + R(\phi_0)\right) \leq \ln\left(\frac{1 + R(\phi)}{1 + R(\phi_0)}\right) \leq \kappa \int_{\phi_0}^{\phi} \left[\sqrt{6} - \alpha(\phi)\right] d\phi.
\ee
Therefore, for 
$\alpha(\phi) > \sqrt{6}$
we get an upper bound on $\phi$,
\be \phi - \phi_0 \leq \max_{\phi'\in [\phi_0,\infty)} \left[\alpha(\phi')-\sqrt{6}\right]^{-1} \kappa^{-1} \ln \left(1+\frac{1}{2}\frac{\dot{\phi}_0^2}{V_0}\right),\ee
implying that the potential can eventually overcome even an arbitrarily large initial kinetic energy and cause the dilaton to stop and turn back towards the weak-coupling regime.

For a perturbative potential which can be expanded at loop orders, each term scales as,
\be V_{n} \sim g_s^{2(n+1)} \sim e^{\sqrt{2}(n+1)\kappa \phi}, \ee
which has an even greater slope, 
\be \alpha^{(n)} = \sqrt{2}(n+1) > \sqrt{6}.\ee
From this, we see that a generic perturbative potential in which there is at least one positive coefficient appearing at $3^{rd}$ loop order or higher is sufficient to cause the dilaton to stop and return towards the weak-coupling regime. 
 The efficiency of this type of stopping potential in a worked example is demonstrated in Fig. \ref{fig:sp_test} in the Appendix.

\section{Summary and Worked Example}
\label{discussion}

This paper's main goal was to illustrate that a cyclic universe can be realized in a scenario in which a combination of string-theoretic ingredients and ideas plays a central role. These include dilaton-gravity, enhanced symmetry points, and, most significantly,  instant folded strings.  Despite the linkages to strings, the evolution is described at every stage by classical equations of motion to leading order, which ensures that the scenario avoids quantum runaway, a multiverse, or other features that prevent it from being conventionally predictive. 
For example, the model predicts a homogeneous and isotropic universe;  no tensor modes (and, hence, no primordial $B$-mode polarization); and time-varying dark energy that leads to an end of accelerated expansion followed by an end to expansion and a transition to slow contraction \cite{Ijjas:2018qbo,Cook:2020oaj}.

Furthermore, the scenario does not rely on non-perturbative aspects of the dilaton potential.   The coefficients determining the perturbative potential satisfy the conditions  $c_3/|c_2|,~\gamma c_2/c_1\gg 1$. These ratios need not be especially large, though, so our working assumption that they are satisfied somewhere in the vast string landscape seems plausible. 

A notable aspect of the model is how the different components of the model work together in such a way that the sequence of stages flows smoothly from one to the other in the necessary order, which is non-trivial because cyclic models require more distinct stages than models assuming a big bang followed only by phases of expansion. 
A prime example is the way the IFSs play a dual role in this construction, which was not anticipated. It was expected that they would mediate a bounce because of their  NEC-violating properties that can arise naturally during contraction and in a way that is under good theoretical control.  But, then, surprisingly, at a much later stage, they turned out to induce dark energy and the current phase of accelerated expansion with the distinctive feature that the equation of state for dark energy is  $w(z)<-1$ for a period before it crosses over to $w(z) >-1$, as detailed at the end of Section~6.  

We close with a fully-worked example of a stable cyclically evolving universe based on all the ingredients and stages described in the previous sections, combined with a choice of plausible parameters adjusted to ensure consistency with current observations. The procedure for selecting parameters used for this example is explained in detail in Appendix~A.  Here, as throughout this paper, we focus on the cosmological background solution.  Generating a nearly scale-invariant spectrum of cosmological density perturbations on top of the background that is consistent with observations and that does not affect the background solution is straightforward, as noted in Section~7.  For example, the mechanism described in \cite{Levy:2015awa,Ijjas:2021ewd}  
in which a second scalar field is added that is kinetically coupled to the dilaton $\phi$ can be adapted without affecting the background solution. Notably, no tensor modes are generated in the process.

\begin{figure}[t!]
\includegraphics[width=15cm]{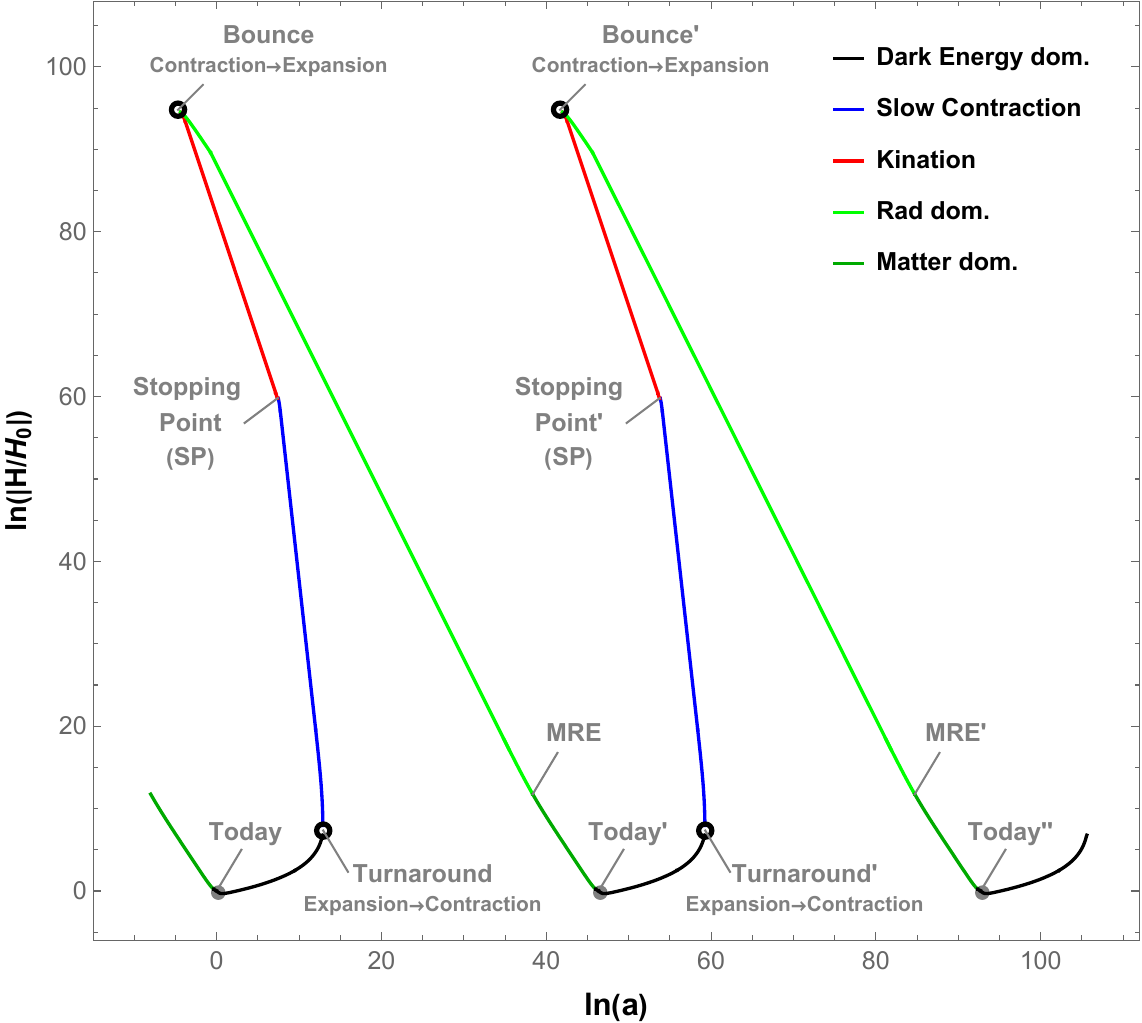}
\centering
\caption{\small A plot showing two consecutive cycles of evolution as a function of $\ln(H(t))$ and $\ln(a(t))$.  Each cycle entails six stages as described in the text; see also Fig.~4.  For the purposes of illustration, the bounce,  the shortest stage which only lasts for a tiny fraction of a second, is represented as an open circle (near the topmost tips). At the bounce, the scale factor $a(t)$ changes from contracting (moving to the left) to expanding (moving to the right). At the complementary point, the turnaround (marked by an open circle as well), the scale factor changes from expanding (moving to the right) to contraction (moving to the left). The Hubble parameter varies periodically, but during each cycle, there is more expansion than contraction.
}
\label{fig:log-log cyclic}
\end{figure}

The cyclic evolution is shown in Figure~\ref{fig:log-log cyclic},  
a  plot of $\ln(|H|)$ versus $\ln(a)$ over two full cycles of $H(t)$. The peculiar shape of the plot is due to a characteristic feature of cyclic bouncing models with periods of slow contraction: namely,   $|H(t)|$ varies periodically but $a(t)$ does not \cite{Ijjas:2019pyf}. Instead, $a(t)$ increases exponentially during periods of expansion and shrinks comparatively little during periods of contraction, such that $a(t)$ increases by a substantial factor from cycle to cycle ($\sim e^{45}$ in this example).
This  exponential
increase in $a(t)$ from one cycle to the next plays a crucial role in diluting the large entropy produced in preceding cycles to a negligible level, and ensuring that the entropy within the Hubble radius is the same after every bounce and purely due to entropy produced by reheating \cite{Ijjas:2021zwv}.

Each cycle can be broken into six stages beginning with ``today'': (1) dark energy-dominated expansion epoch (black); (2) slow contraction epoch (blue); (3) kination contraction epoch (red); (4) bounce (small open circle); (5) radiation dominated expansion (bright green); and, (6) matter-radiation equality and matter dominated expansion epoch (dark green).

\vspace{6mm}
\noindent
{\Large{\bf Acknowledgements}}

\noindent
We thank Alek Bedroya for his helpful comments.
NI and UP are supported in part by the ISF through grant number 256/22. 
PJS  is supported in part by the DOE through grant number DEFG02-91ER40671
and by the Simons Foundation through grant number 654561. 

\appendix

\section{Constructing a working example}
\label{The worked example in more detail}

In this Appendix, we describe the procedure for choosing parameters in accordance with the constraints outlined in the preceding sections of the paper, as was  
used in constructing the worked example shown in Fig. \ref{fig:log-log cyclic}.  In some cases, there are no significant physical constraints, so our choice was made to attain a more aesthetic, well-proportioned plot. 

\begin{itemize}
\item \textbf{The potential:} First, we must specify the form of the potential. For simplicity, we have chosen to focus on a perturbative potential of the form
\be V(\phi) \equiv -c_2 g_s^6 + c_3 g_s^8.\ee
The 1-loop term, proportional to $g_s^4$, has been omitted here for simplicity since it plays a much less significant role in the modification of the effective potential as mentioned in Section \ref{nutshell} and demonstrated in Section \ref{LambdaCDM}, but the generalization to any potential of the form described in Section \ref{LambdaCDM} with $\mu\gg 1$ is straightforward.

The ratio of $c_3$ and $c_2$  determines the maximal value of $g_s$, which is the value reached at the stopping point SP in Fig.~4. More precisely, one finds that
\be g_s |_{\text{max}} \equiv g_{\scriptscriptstyle{SP}} \approx
\sqrt{\frac{c_2}{c_3}}.\ee
The fact that the stopping point is reached almost immediately after the potential becomes positive is a testament to the efficiency of the stopping potential. For more details, see Fig. \ref{fig:sp_test}, which shows this numerically.

In our example in Fig.~\ref{fig:log-log cyclic}, we have set,
\be g_s |_{\text{max}} \equiv g_{\scriptscriptstyle{SP}} \approx
\sqrt{\frac{c_2}{c_3}} = 0.1, \ee
sufficiently small to avoid the strong coupling region and maintain perturbative control.

We specify $c_2$ only after identifying the value of $g_s$ ``today,'' so that our expression for the effective potential  $V_{eff}$ agrees with the dark-energy density observed today.
\item \textbf{The enhanced symmetry point:} The most important parameter to consider at the ESP is the value of $\phi_{\scriptscriptstyle{ESP}}$, and the corresponding value of  $g_s \equiv g_{\scriptscriptstyle{ESP}}$.
Generically, so long as the parameters chosen are not too close to saturating the bound of (\ref{ESP bound}), the minimum value of $g_s$ during a cycle is
\be g_s|_{\text{min}} \approx g_{\scriptscriptstyle{ESP}}.\ee

The temperature right after the bounce, the reheat temperature $T_{\rm rh}$, is set approximately by the maximum value of $(H M_{\rm Pl})^{1/2}$, which in turn is related to the evolution of $g_s$ during the cycle. Below, we show that the reheat temperature can be related to $g_s|_{\rm max}$ and $g_s|_{\rm min}$, namely
\be
\ln\left( \frac{T_{rh}}{T_0}\right)  \cong \frac{1}{2}\left(3+\sqrt{3}\right)\ln \left(\frac{g_s|_{\rm max}}{g_s|_{\rm min}}\right) + \mathcal{O}(1),
\label{rhratio}
\ee
where $T_0 \approx 2.3 \times 10^{-13}$ GeV the temperature today.

In our example in Fig.~\ref{fig:log-log cyclic}, we make the choice of
\be 
g_s|_{\text{max}} =0.1 \; \; {\rm and} \; \;  
g_s|_{\text{min}} \approx g_{\scriptscriptstyle{ESP}} = 10^{-10}, \label{g min}\ee
which implies from \ref{rhratio} a re-heating temperature of
\be T_{rh} \sim 10^{8}~\text{Gev}. \label{T re}\ee
Note that our values for (\ref{g min}) and (\ref{T re}) are consistent with (\ref{yg}) as well as the condition $T_{rh} \ll M_s\equiv g_s M_{\text{Pl}}$, which is required to ensure the validity of our low-energy effective description, and avoid the thermalization of massive string states.

The ESP also has two more ``secondary'' parameters associated with it. These are the coupling strength $\lambda_{\chi}$, and the typical life-time of the $\chi$ particles before they eventually decay into standard model particles during reheating, let's label it $\tau_\chi$.

One has to ensure that the coupling $\lambda_{\chi}$ associated with the ESP is large enough so that our value of $\kappa\dot{\phi}_{\scriptscriptstyle{ESP}}\sim H_{\scriptscriptstyle{ESP}}\sim M_{\text{Pl}}^{-1}\,T_{rh}^2$ is consistent with the bound set by (\ref{fc}). In our example, we have set 
\be \lambda_{\chi} = 10^{-3},\ee
which results in a value for $\dot{\phi}$ that is easily (by two orders of magnitude) below the bound in (\ref{fc}).

The life-time associated with the decay of the ESP particles $\tau_\chi$, has to be substantially longer than the stopping time $t_{*}$ as defined in (\ref{stopping-time}) --- so that there is ample time for the $\chi$ particles to stop the dilaton and generate a bounce before undergoing decay.

For the purpose of generating Fig. \ref{fig:log-log cyclic}, we have arbitrarily set
\be \tau_{\chi} = 100 \, t_*.\ee

\item \textbf{IFS decay:} The only remaining parameter in our model is the dimensionless constant $\gamma$ that is related to the life-time of IFS. 
One constraint on $\gamma$ comes from the analysis in Section \ref{LambdaCDM} that checks the consistency of the IFS-induced dark-energy phase in the cyclic picture with observations; see (\ref{rough limits}) and the associated discussion for more information.

Since as discussed below, the value of $g_s$ today is fairly close to the minimal value within a cycle $g_s |_{\text{min}}$, therefore (\ref{rough limits}) effectively sets an upper-bound on $\gamma^{-1}g_s^2 |_{\text{min}}$ or, equivalently, a lower bound on $\gamma$.

This is not a very tight constraint.  For the purposes of making a well-proportioned plot in Fig.~\ref{fig:log-log cyclic},  we have set
\be \gamma = 5 \times 10^3 \, g_s^2 |_{\text{min}} = 5 \times 10^{-17}, \label{gammab}\ee
to be about as low as possible.
Larger values of $\gamma$ lead to a longer dark energy-dominated phase that stretches the figure in a way that makes it difficult to visually resolve the other phases.

More generally, an upper bound on $\gamma$ can be obtained from the requirement that the number of  $e$-foldings of $H$ during the dark energy-dominated expansion phase (as estimated in (\ref{H estimate}))  does not exceed  40 or so, allowing at least   60 $e$-foldings during contraction  (both the slow contraction and kination stages) in order to be able to generate the nearly scale invariant spectrum of density fluctuations, {\it e.g.,} using the mechanism described in \cite{Levy:2015awa,Ijjas:2021ewd}. Using (\ref{H estimate}) to estimate the number of $H$ $e$-foldings, and requiring that it is less than 40, we obtain the bound
\be \gamma \lesssim 2\times 10^{-7}, \label{upper gamma}\ee which means that it is possible to increase $\gamma$ such that the dark energy-dominated phase lasts much longer than in our worked example.

It is important to emphasize that the bound in (\ref{upper gamma}) is not universal but rather specific to the parameter choices used in our example. Values of $\gamma \gtrsim 10^{-7}$ can still be used to construct viable cyclic models that are under good perturbative control. For instance, one may adopt a steeper (higher-order) perturbative potential, which reduces the required range of $g_s$ to achieve the same number of $H$ $e$-foldings. Alternatively, lowering the reheating temperature decreases the total number of $e$-foldings needed, thereby also relaxing the constraint on $\gamma$.

Smoothing and flattening of space does not impose an additional constraint.  Between the dark energy-dominated and slow contraction phases, there is more than sufficient smoothing and flattening to solve the homogeneity, isotropy, and flatness problems if the conditions for generating the nearly scale-invariant spectrum of density fluctuations are satisfied.

\item \textbf{Reheating considerations:} 
To estimate the reheating temperature $T_{rh}$, we track the change in $H$ and the temperature from their present-day values (with $T_0 \sim 2.3 \times 10^{-13}$ GeV) to their maximal values at reheating just after the bounce.

At the onset of the dark-energy-dominated phase, using (\ref{DE density}), 
\be  3H_0^2 \cong \kappa^2 \rho_{DE} \cong \frac{1}{2}\kappa ^2c_2 \, g_s^6 \sim c_2 g_s^6. \ee
During the slow contraction phase, a simple estimate yields
\be 3H^2 \cong \frac{1}{1-\varepsilon} \, \kappa^2 V \cong \frac{1}{8} \kappa^2 c_2 \, g_s^6 \sim c_2 g_s^6.\ee
The efficiency of the stopping mechanism is illustrated in Figure~\ref{fig:sp_test}, which 
shows that the stopping period leads to a negligible change (less than $\mathcal{O}(1)$) in both $\ln(H)$ and $\ln(g_s^2)$.

\begin{figure}[!t]
\includegraphics[width=11cm]{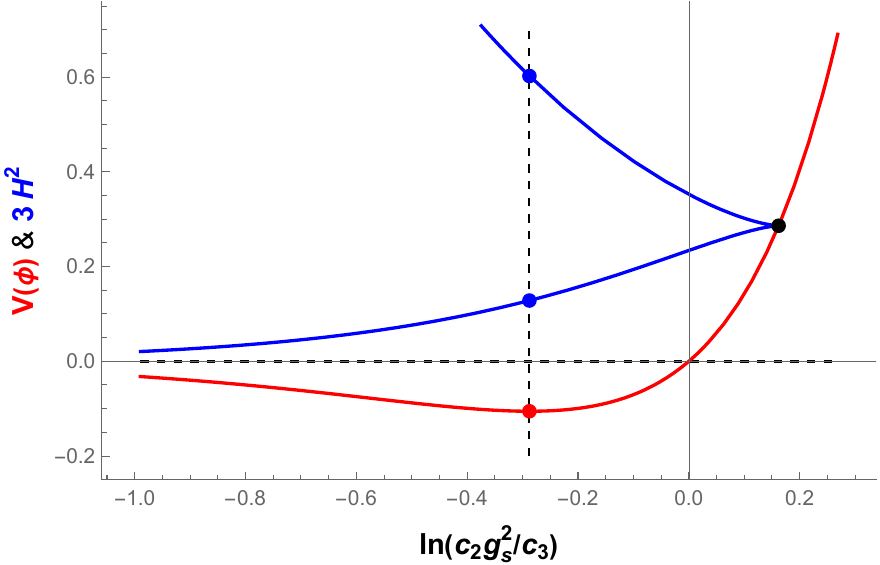}
\centering
\caption{Efficiency of the stopping potential. The red curve depicts the potential whose approximate form is $V(\phi) = -c_2 g_s^6 + c_3 g_s^8$, while the blue curve shows the value of $3H^2 = V + \frac{1}{2}\dot{\phi}^2$. By definition, $3H^2$ is positive and never less than $V$.  The two curves intersect only at the stopping point (SP), where $\dot{\phi} = 0$.
A  test of the stopping mechanism's efficiency is the number of  $e$-foldings of $H$ that occur between the first crossing of the minimum (as the dilaton rolls toward strong coupling) and the second crossing (as it rebounds and rolls toward weak coupling). In our case, this number is small: $\mathcal{R}_{SP} \equiv \ln(H_2/H_1) \approx 0.77$, indicating that the stopping potential is quite efficient since only a negligible number of   $e$-foldings of $H$ occur between these crossings. This also implies that $g_{\scriptscriptstyle{SP}} \approx
\sqrt{c_2/c_3}$.}
\label{fig:sp_test}
\end{figure}

From this, we conclude that the number of $e$-foldings of $H(t)$ between today and the stopping point (SP) is 
\be \ln\left( \frac{H_{SP}}{H_0}\right) \cong 3\ln \left(\frac{g_{\scriptscriptstyle{SP}}}{g_s|_{\rm today}}\right) + \mathcal{O}(1). \label{eb-1}\ee
After reaching the SP, the universe enters a kination phase, lasting until the ESP is reached and the bounce is triggered. During this phase, $H$ scales with the dilaton as
\be  \ln\left( \frac{H_{ESP}}{H_{SP}}\right) \cong \sqrt{3}\ln \left(\frac{g_{\scriptscriptstyle{SP}}}{g_{\scriptscriptstyle{ESP}}}\right) + \mathcal{O}(1). \label{eb-2}\ee
Nearly the entire evolution from the bounce until today occurs during the radiation- and matter-dominated phases when the dilaton is effectively frozen by Hubble friction. Furthermore, in the relatively short interval between when the dilaton becomes unfrozen and today, the evolution is the IFS slow-roll limit.  Hence,  there is only a small change in $g_s$;  see the discussion in Section \ref{nutshell} for more detail.)
Consequently, the values of $g_s$ today, at its minimum, and at the ESP are fairly close
\be g_s|_{\rm today} \approx g_s|_{\rm min} \approx g_{\scriptscriptstyle{ESP}}. \label{eb-3}\ee
Combining (\ref{eb-1}), (\ref{eb-2}), and (\ref{eb-3}) yields:
\be \ln\left( \frac{T_{rh}}{T_0}\right) \cong \frac{1}{2}\ln\left( \frac{H_{ESP}}{H_{0}}\right) \cong \frac{1}{2}\left(3+\sqrt{3}\right)\ln \left(\frac{g_s|_{\rm max}}{g_s|_{\rm min}}\right) + \mathcal{O}(1). \ee

Note that this relation applies specifically to the two-loop potential we have chosen. If instead the dynamics of the dilaton were driven by a non-perturbative potential of the form
\be V(\phi) \cong -C e^{\frac{\phi}{M}}, \quad \epsilon \equiv \frac{1}{2}\left(\frac{M_{\text{Pl}}}{M}\right)^2,\ee
as in many examples of slow-contraction, a similar argument would have led to the more general relation
\be
\ln\left( \frac{T_{rh}}{T_0}\right)  \cong \frac{1}{2}\left(\sqrt{\epsilon}+\sqrt{3}\right)\ln \left(\frac{g_s|_{\rm max}}{g_s|_{\rm min}}\right) + \mathcal{O}(1),
\ee
As a result, for a steeper potential and larger $\epsilon$ during slow contraction (while keeping $T_{rh}$ fixed), the constraints we have derived based on $g_s|_{\rm min}$ are greatly relaxed, and what is already a large range of acceptable parameters becomes much larger.
\end{itemize}

\bibliographystyle{ieeetr}
\bibliography{references}

\begin{thebibliography}{10}

\bibitem{Penrose:1988mg}
R.~Penrose, ``{Difficulties with inflationary cosmology},'' {\em A.ls N.Y.Acad.Sci.}, vol.~571, pp.~249--264, 1989.

\bibitem{Hartle:1983ai}
J.~B. Hartle and S.~W. Hawking, ``Wave function of the universe,'' {\em Phys. Rev. D}, vol.~28, pp.~2960--2975, Dec 1983.

\bibitem{carroll2010}
S.~M. Carroll, {\em From Eternity to Here: The Quest for the Ultimate Theory of Time}.
\newblock New York: Plume, 2010.

\bibitem{Carroll:2010aj}
S.~M. Carroll and H.~Tam, ``{Unitary Evolution and Cosmological Fine-Tuning},'' 7 2010.

\bibitem{Garfinkle:2023vzf}
D.~Garfinkle, A.~Ijjas, and P.~J. Steinhardt, ``{Initial conditions problem in cosmological inflation revisited},'' {\em Phys. Lett. B}, vol.~843, p.~138028, 2023.

\bibitem{Ijjas2024}
A.~Ijjas, P.~J. Steinhardt, D.~Garfinkle, and W.~G. Cook, ``Smoothing and flattening the universe through slow contraction versus inflation,'' {\em Journal of Cosmology and Astroparticle Physics}, vol.~2024, no.~07, p.~077, 2024.

\bibitem{Linde:1981mu}
A.~D. Linde, ``{A New Inflationary Universe Scenario: A Possible Solution of the Horizon, Flatness, Homogeneity, Isotropy and Primordial Monopole Problems},'' {\em Phys.Lett.}, vol.~B108, pp.~389--393, 1982.

\bibitem{Albrecht:1982wi}
A.~Albrecht and P.~J. Steinhardt, ``Cosmology for grand unified theories with radiatively induced symmetry breaking,'' {\em Phys.Rev.Lett.}, vol.~48, pp.~1220--1223, 1982.

\bibitem{Steinhardt:1984jj}
P.~Steinhardt and M.~S. Turner, ``{A Prescription for Successful New Inflation},'' {\em Phys.Rev.}, vol.~D29, pp.~2162--2171, 1984.

\bibitem{Steinhardt1983}
P.~J. Steinhardt, ``Natural inflation,'' in {\em The Very Early Universe} (G.~W. Gibbons, S.~W. Hawking, and S.~T.~C. Siklos, eds.), pp.~251--266, Cambridge, UK: Cambridge University Press, 1983.

\bibitem{Vilenkin:1983xq}
A.~Vilenkin, ``{The Birth of Inflationary Universes},'' {\em Phys.Rev.}, vol.~D27, p.~2848, 1983.

\bibitem{Linde1986}
A.~D. Linde, ``Eternal chaotic inflation,'' {\em Modern Physics Letters A}, vol.~1, no.~2, pp.~81--85, 1986.

\bibitem{Linde:1986fd}
A.~D. Linde, ``{Eternally Existing Self-reproducing Chaotic Inflationary Universe},'' {\em Phys. Lett.}, vol.~B175, pp.~395--400, 1986.

\bibitem{Garriga:2005av}
J.~Garriga, D.~Schwartz-Perlov, A.~Vilenkin, and S.~Winitzki, ``{Probabilities in the inflationary multiverse},'' {\em JCAP}, vol.~0601, p.~017, 2006.

\bibitem{Carr2007}
B.~Carr, ed., {\em Universe or Multiverse?}
\newblock Cambridge: Cambridge University Press, 2007.

\bibitem{Guth:2007ng}
A.~H. Guth, ``{Eternal inflation and its implications},'' {\em J. Phys.}, vol.~A40, pp.~6811--6826, 2007.

\bibitem{Garriga2008}
J.~Garriga and A.~Vilenkin, ``Prediction and explanation in the multiverse,'' {\em Physical Review D}, vol.~77, p.~043526, 2008.

\bibitem{Linde2015}
A.~Linde, ``A brief history of the multiverse,'' {\em Reports on Progress in Physics}, 2015.

\bibitem{Nomura2017}
Y.~Nomura, ``The quantum multiverse,'' {\em Scientific American}, vol.~316, no.~3, pp.~28--35, 2017.

\bibitem{Abbott1984}
L.~F. Abbott and M.~B. Wise, ``Constraints on generalized inflationary cosmologies,'' {\em Nuclear Physics B}, vol.~244, pp.~541--548, 1984.

\bibitem{Starobinsky1983}
A.~A. Starobinsky, ``The perturbation spectrum evolving from a quantum fluctuation in a de sitter space,'' {\em JETP Letters}, vol.~30, pp.~682--685, 1983.
\newblock Pisma Zh. Eksp. Teor. Fiz. 30, 719–723 (1983).

\bibitem{Flauger:2014qra}
R.~Flauger, J.~C. Hill, and D.~N. Spergel, ``Toward an understanding of foreground emission in the bicep2 region,'' {\em Journal of Cosmology and Astroparticle Physics}, vol.~2014, no.~08, p.~039, 2014.

\bibitem{BICEP2Planck2015}
BICEP2/Keck and P.~Collaborations, ``A joint analysis of bicep2/keck array and planck data,'' {\em Physical Review Letters}, vol.~114, p.~101301, 2015.

\bibitem{BICEP2021}
B.~Collaboration, ``Improved constraints on primordial gravitational waves using planck, wmap, and bicep/keck observations through the 2018 observing season,'' {\em Physical Review Letters}, vol.~127, p.~151301, 2021.

\bibitem{MartinBrandenberger2001}
J.~Martin and R.~H. Brandenberger, ``Trans-planckian issues for inflationary cosmology,'' {\em Physical Review D}, vol.~63, p.~123501, 2001.

\bibitem{Agrawal:2018own}
P.~Agrawal, G.~Obied, P.~J. Steinhardt, and C.~Vafa, ``{On the Cosmological Implications of the String Swampland},'' {\em Phys. Lett.}, vol.~B784, pp.~271--276, 2018.

\bibitem{Palti2019}
E.~Palti, ``The swampland: Introduction and review,'' {\em Fortschritte der Physik}, vol.~67, no.~6, p.~1900037, 2019.

\bibitem{Bedroya:2019snp}
A.~Bedroya and C.~Vafa, ``{Trans-Planckian Censorship and the Swampland},'' {\em JHEP}, vol.~09, p.~123, 2020.

\bibitem{Brandenberger2021}
R.~Brandenberger, ``Trans-planckian censorship conjecture and early universe cosmology,'' {\em International Journal of Modern Physics D}, vol.~30, no.~14, p.~2140004, 2021.

\bibitem{Bedroya2022}
A.~Bedroya, ``Holographic origin of tcc and the distance conjecture,'' {\em arXiv preprint}, 2022.

\bibitem{Bedroya2024}
A.~Bedroya, Q.~Lu, and P.~J. Steinhardt, ``Tcc in the interior of moduli space and its implications for the string landscape and cosmology,'' {\em arXiv preprint}, 2024.

\bibitem{Das:2018rpg}
S.~Das, ``{Warm Inflation in the light of Swampland Criteria},'' {\em Phys. Rev. D}, vol.~99, no.~6, p.~063514, 2019.

\bibitem{Das:2019hto}
S.~Das, ``{Distance, de Sitter and Trans-Planckian Censorship conjectures: the status quo of Warm Inflation},'' {\em Phys. Dark Univ.}, vol.~27, p.~100432, 2020.

\bibitem{Brandenberger:2019eni}
R.~Brandenberger and E.~Wilson-Ewing, ``{Strengthening the TCC Bound on Inflationary Cosmology},'' {\em JCAP}, vol.~03, p.~047, 2020.

\bibitem{Brandenberger:2020oav}
R.~Brandenberger, V.~Kamali, and R.~O. Ramos, ``{Strengthening the de Sitter swampland conjecture in warm inflation},'' {\em JHEP}, vol.~08, p.~127, 2020.

\bibitem{Vafa2022}
N.~B. Agmon, A.~Bedroya, M.~J. Kang, and C.~Vafa, ``Lectures on the string landscape and the swampland,'' {\em arXiv preprint}, 2022.

\bibitem{Ijjas:2018qbo}
A.~Ijjas and P.~J. Steinhardt, ``{Bouncing Cosmology made simple},'' {\em Class. Quant. Grav.}, vol.~35, no.~13, p.~135004, 2018.

\bibitem{Cook:2020oaj}
W.~G. Cook, I.~A. Glushchenko, A.~Ijjas, F.~Pretorius, and P.~J. Steinhardt, ``{Supersmoothing through Slow Contraction},'' {\em Phys. Lett. B}, vol.~808, p.~135690, 2020.

\bibitem{Ijjas:2020dws}
A.~Ijjas, W.~G. Cook, F.~Pretorius, P.~J. Steinhardt, and E.~Y. Davies, ``{Robustness of slow contraction to cosmic initial conditions},'' {\em JCAP}, vol.~08, p.~030, 2020.

\bibitem{Ijjas:2021zwv}
A.~Ijjas and P.~J. Steinhardt, ``{Entropy, black holes, and the new cyclic universe},'' {\em Phys. Lett. B}, vol.~824, p.~136823, 2022.

\bibitem{Ijjas:2021gkf}
A.~Ijjas, A.~P. Sullivan, F.~Pretorius, P.~J. Steinhardt, and W.~G. Cook, ``{Ultralocality and slow contraction},'' {\em JCAP}, vol.~06, p.~013, 2021.

\bibitem{Kist:2022mew}
T.~Kist and A.~Ijjas, ``{The robustness of slow contraction and the shape of the scalar field potential},'' {\em JCAP}, vol.~08, no.~08, p.~046, 2022.

\bibitem{Steinhardt:2002ih}
P.~J. Steinhardt, N.~Turok, and N.~Turok, ``{A Cyclic model of the universe},'' {\em Science}, vol.~296, pp.~1436--1439, 2002.

\bibitem{Ijjas:2019pyf}
A.~Ijjas and P.~J. Steinhardt, ``{A new kind of cyclic universe},'' {\em Phys. Lett.}, vol.~B795, pp.~666--672, 2019.

\bibitem{Montefalcone:2020vlu}
G.~Montefalcone, P.~J. Steinhardt, and D.~H. Wesley, ``{Dark energy, extra dimensions, and the Swampland},'' {\em JHEP}, vol.~06, p.~091, 2020.

\bibitem{Andrei:2022rhi}
C.~Andrei, A.~Ijjas, and P.~J. Steinhardt, ``{Rapidly descending dark energy and the end of cosmic expansion},'' {\em Proc. Nat. Acad. Sci.}, vol.~119, no.~15, p.~e2200539119, 2022.

\bibitem{DESI:2024mwx}
A.~G. Adame {\em et~al.}, ``{DESI 2024 VI: cosmological constraints from the measurements of baryon acoustic oscillations},'' {\em JCAP}, vol.~02, p.~021, 2025.

\bibitem{DESI:2025fii}
K.~Lodha {\em et~al.}, ``{Extended Dark Energy analysis using DESI DR2 BAO measurements},'' 3 2025.

\bibitem{Brandenberger:1988aj}
R.~H. Brandenberger and C.~Vafa, ``{Superstrings in the Early Universe},'' {\em Nucl. Phys. B}, vol.~316, pp.~391--410, 1989.

\bibitem{Khoury:2001wf}
J.~Khoury, B.~A. Ovrut, P.~J. Steinhardt, and N.~Turok, ``{The Ekpyrotic universe: Colliding branes and the origin of the hot big bang},'' {\em Phys. Rev. D}, vol.~64, p.~123522, 2001.

\bibitem{Khoury:2001bz}
J.~Khoury, B.~A. Ovrut, N.~Seiberg, P.~J. Steinhardt, and N.~Turok, ``{From big crunch to big bang},'' {\em Phys. Rev. D}, vol.~65, p.~086007, 2002.

\bibitem{Lehners:2011kr}
J.-L. Lehners, ``{Cosmic Bounces and Cyclic Universes},'' {\em Class.Quant.Grav.}, vol.~28, p.~204004, 2011.

\bibitem{Easson:2011zy}
D.~A. Easson, I.~Sawicki, and A.~Vikman, ``{G-Bounce},'' {\em JCAP}, vol.~1111, p.~021, 2011.

\bibitem{Xue2013}
B.~Xue, D.~Garfinkle, F.~Pretorius, and P.~J. Steinhardt, ``Nonperturbative analysis of the evolution of cosmological perturbations through a nonsingular bounce,'' {\em Physical Review D}, vol.~88, no.~8, p.~083509, 2013.

\bibitem{Qiu:2013eoa}
T.~Qiu, X.~Gao, and E.~N. Saridakis, ``{Towards Anisotropy-Free and Non-Singular Bounce Cosmology with Scale-invariant Perturbations},'' {\em Phys.Rev.}, vol.~D88, p.~043525, 2013.

\bibitem{Battarra:2014tga}
L.~Battarra, M.~Koehn, J.-L. Lehners, and B.~A. Ovrut, ``{Cosmological Perturbations Through a Non-Singular Ghost-Condensate/Galileon Bounce},'' {\em JCAP}, vol.~1407, p.~007, 2014.

\bibitem{Alexander:2014uaa}
S.~Alexander, Y.-F. Cai, and A.~Marciano, ``{Fermi-bounce cosmology and the fermion curvaton mechanism},'' {\em Phys. Lett. B}, vol.~745, pp.~97--104, 2015.

\bibitem{Ijjas:2016tpn}
A.~Ijjas and P.~J. Steinhardt, ``{Classically stable nonsingular cosmological bounces},'' {\em Phys. Rev. Lett.}, vol.~117, no.~12, p.~121304, 2016.

\bibitem{Ijjas:2016vtq}
A.~Ijjas and P.~J. Steinhardt, ``{Fully stable cosmological solutions with a non-singular classical bounce},'' {\em Phys. Lett. B}, vol.~764, pp.~289--294, 2017.

\bibitem{Ijjas:2017pei}
A.~Ijjas, ``{Space-time slicing in Horndeski theories and its implications for non-singular bouncing solutions},'' {\em JCAP}, vol.~1802, no.~02, p.~007, 2018.

\bibitem{Farnsworth:2017wzr}
S.~Farnsworth, J.-L. Lehners, and T.~Qiu, ``{Spinor driven cosmic bounces and their cosmological perturbations},'' {\em Phys. Rev.}, vol.~D96, no.~8, p.~083530, 2017.

\bibitem{Brandenberger:2016vhg}
R.~Brandenberger and P.~P., ``{Bouncing Cosmologies: Progress and Problems},'' {\em Found. Phys.}, vol.~47, no.~6, pp.~797--850, 2017.

\bibitem{Tukhashvili:2023itb}
G.~Tukhashvili and P.~J. Steinhardt, ``{Cosmological Bounces Induced by a Fermion Condensate},'' {\em Phys. Rev. Lett.}, vol.~131, no.~9, p.~091001, 2023.

\bibitem{InstantCosmology}
N.~Itzhaki and U.~Peleg, ``{Instant Cosmology},'' {\em {Journal of High Energy Physics}}, 12 2024.

\bibitem{Itzhaki:2018glf}
N.~Itzhaki, ``Stringy instability inside the black hole,'' {\em Journal of High Energy Physics}, vol.~2018, 2018.

\bibitem{ESP}
L.~Kofman, A.~Linde, X.~Liu, A.~Maloney, L.~Mcallister, and E.~Silverstein, ``Beauty is attractive: Moduli trapping at enhanced symmetry points,'' {\em Journal of High Energy Physics}, vol.~2004, p.~30, may 2004.

\bibitem{Dine:1985he}
M.~Dine and N.~Seiberg, ``{Is the Superstring Weakly Coupled?},'' {\em Phys. Lett. B}, vol.~162, pp.~299--302, 1985.

\bibitem{Attali:2018goq}
K.~Attali and N.~Itzhaki, ``{The Averaged Null Energy Condition and the Black Hole Interior in String Theory},'' {\em Nucl. Phys. B}, vol.~943, p.~114631, 2019.

\bibitem{Itzhaki:2021scf}
N.~Itzhaki, ``{String Theory and The Arrow of Time},'' {\em JHEP}, vol.~03, p.~192, 2021.

\bibitem{Hashimoto:2022dro}
A.~Hashimoto, N.~Itzhaki, and U.~Peleg, ``{A worldsheet description of instant folded strings},'' {\em JHEP}, vol.~02, p.~088, 2023.

\bibitem{Creminelli:2007aq}
P.~Creminelli and L.~Senatore, ``{A smooth bouncing cosmology with scale invariant spectrum},'' {\em JCAP}, vol.~0711, p.~010, 2007.

\bibitem{Levy:2015awa}
A.~M. Levy, A.~Ijjas, and P.~J. Steinhardt, ``{Scale-invariant perturbations in ekpyrotic cosmologies without fine-tuning of initial conditions},'' {\em Phys. Rev. D}, vol.~92, no.~6, p.~063524, 2015.

\bibitem{Brandenberger:2020ekpyrotic}
R.~H. Brandenberger and Z.~Wang, ``{Nonsingular ekpyrotic cosmology with a nearly scale-invariant spectrum of cosmological perturbations and gravitational waves},'' {\em Phys. Rev. D}, vol.~101, no.~6, p.~063522, 2020.
\newblock Published March 20, 2020.

\bibitem{Ijjas:2021ewd}
A.~Ijjas and R.~Kolevatov, ``{Nearly scale-invariant curvature modes from entropy perturbations during the graceful exit phase},'' {\em Phys. Rev. D}, vol.~103, no.~10, p.~L101302, 2021.

\end{thebibliography}
\end{document}